\documentclass[12pt]{article}
\usepackage{graphicx}
\usepackage{amsmath}
\usepackage{mathrsfs}
\newcommand{\eps}{\varepsilon}
\def\beq{\begin{equation}}
\def\eeq{\end{equation}}
\def\bea{\begin{eqnarray}}
\def\eea{\end{eqnarray}}
\def\nn{\nonumber}
\def\nl{\nonumber \\}
\def\roughly#1{\mathrel{\raise.3ex\hbox
{$#1$\kern-.75em\lower1ex\hbox{$\sim$}}}}
\def\lsim{\roughly<}
\def\gsim{\roughly>}
\def\sss{\scriptscriptstyle}

\def\bq{B_q^0}

\def\bra#1{\left\langle  #1\right|} \def\ket#1{\left| #1\right\rangle}
\def\barpk{{\raise.35ex\hbox  {${\sss  (}$}}--{\raise.35ex\hbox{${\sss
)}$}}}        \def\bbarp{\hbox{$B$\kern-0.9em\raise1.4ex\hbox{\barpk}}}

\def\lsim{\roughly<}  \def\gsim{\roughly>}

\def\bra#1{\left\langle  #1\right|} \def\ket#1{\left| #1\right\rangle}

  \def\rr2{{1\over\sqrt{2}}}

\def\.{\!\cdot\!}    \def\:{\cdots}   \def\[{\left[}   \def\]{\right]}
\def\({\left(} \def\){\right)} 
\usepackage{graphicx}
\usepackage{amsmath}
\usepackage{mathrsfs}
\def\nn{\nonumber}
\def\nl{\nonumber \\}

\def\bq{\begin{eqnarray}}
\def\eq{\end{eqnarray}}
\def\nn{\nonumber}

\usepackage{epsfig}
\usepackage{dcolumn}
\usepackage{bm}

\newcommand{\comment}[1]{}
\newcommand{\newc}{\newcommand}

\newcommand{\be}{\begin{equation}}
\newcommand{\ee}{\end{equation}}
\newcommand{\ba}{\begin{array}}
\newcommand{\ea}{\end{array}}

\newcommand{\dsl}{\hspace{-5.5pt}/}

\def\lsim{\ ^<\llap{$_\sim$}\ }
\def\gsim{\ ^>\llap{$_\sim$}\ }
\def\r2{\sqrt 2}
\def\beq{\begin{equation}}
\def\eeq{\end{equation}}
\def\bea{\begin{eqnarray}}
\def\eea{\end{eqnarray}}
\def\nn{\nonumber}

\def\bra#1{\left\langle #1\right|}
\def\ket#1{\left| #1\right\rangle}

\def\lsim{\mathrel{\mathpalette\@versim<}}
\def\gsim{\mathrel{\mathpalette\@versim>}}
\def\@versim#1#2{\vcenter{\offinterlineskipresembel
    \ialign{$\m@th#1\hfil##\hfil$\crcr#2\crcr\sim\crcr } }}

\newc{\non}{\noindent}

\def\scat{ \nu_{\tau}+ n \to \tau + X}

\def\scatdelta{ \nu_{\tau}+ n \to \tau^- + \Delta^+}
\def\scatdeltaanti{ \bar{\nu}_{\tau}+ p \to \tau^+ + \Delta^0}
\def\scatdeltamix{ \nu_{\tau}(\bar{\nu}_{\tau})+ n(p) \to \tau^-(\tau^+) + \Delta^+(\Delta^0)}
\def\nutau{ \nu_{\tau}}

\def\taud{\tau^- \to \pi^- \nu_{\tau}}
\def\tauv{\tau^- \to \rho^- \nu_{\tau}}

\def\bra{\langle}
\def\ket{\rangle}

\def\bra#1{\left\langle #1\right|}
\def\ket#1{\left| #1\right\rangle}

\def\lsim{\mathrel{\mathpalette\@versim<}}
\def\gsim{\mathrel{\mathpalette\@versim>}}
\def\@versim#1#2{\vcenter{\offinterlineskipresembel
    \ialign{$\m@th#1\hfil##\hfil$\crcr#2\crcr\sim\crcr } }}

\def\EPSFIG[#1]#2#3#4{		
\begin{figure}[H]		
\begin{center}			%
\includegraphics[#1]{#2}	%
\end{center}			%
\caption{#3}			%
\label{#4}			%
\end{figure}			%
}                              %

\def\nutau{ \nu_{\tau}}

\def\taud{\tau^- \to \pi^- \nu_{\tau}}
\def\tauv{\tau^- \to \rho^- \nu_{\tau}}
\def\scat{ \nu_{\tau}+ N \to \tau^- + X}
\def\scatanti{ \bar{\nu}_{\tau}+ N \to \tau^+ + X}
\def\scatmix{ \nu_{\tau}(\bar{\nu}_{\tau})+ N \to \tau^-(\tau^+) + X}

\def\bra{\langle}
\def\ket{\rangle}

\begin{document}

 \unitlength = 1mm
\begin{flushright}
UMISS-HEP-2013-05 \\
[10mm]
\end{flushright}

\begin{center}
\bigskip {\Large  \bf Tau neutrino as a probe of nonstandard interaction}
\\[8mm]
Ahmed Rashed $^{\dag \ddag}$
\footnote{E-mail:
\texttt{amrashed@phy.olemiss.edu}}
, Preet Sharma $^{\dag}$
\footnote{E-mail:
\texttt{preetsharma@phy.olemiss.edu}}
 and Alakabha Datta $^{\dag}$ 
\footnote{E-mail:
\texttt{datta@phy.olemiss.edu}} 
\\[3mm]
\end{center}

\begin{center}
~~~{\it $^{\dag}$ Department  of Physics and Astronomy,}\\ 
~~~{\it University of Mississippi,}\\
~~~{\it Lewis Hall, University, MS, 38677,USA}\\
\end{center}

\begin{center}
~~~{\it $^{\ddag}$ Department  of Physics, Faculty of Science,}\\ 
~~~{\it Ain Shams University, Cairo, 11566, Egypt}\\
\end{center}


\begin{center} 
\bigskip (\today) \vskip0.5cm {\Large Abstract\\} \vskip3truemm
\parbox[t]{\textwidth}{
We study the $\Delta$-resonance and deep inelastic scattering contributions in the tau-neutrino nucleon scattering $\scat$ and $\scatanti$ in the presence of a charged Higgs and a $W'$ gauge boson. The new physics effects to the quasielastic process have been discussed in a previous work. The extractions of the atmospheric  mixing angle $\theta_{23}$ 
rely on the standard model cross sections for $\scat$ in $\nu_{\tau}$ appearance experiments. Corrections to the cross sections from the charged Higgs and $W'$ contributions  modify the measured mixing angle. We include form factor effects in the new physics calculations and find the deviations of the mixing angle. If  high-energy Long Base Line experiments are designed to measure $\theta_{13}$ through tau neutrino appearance, the new physics effects to $\scat$ and $\scatanti$ can impact the extraction of this mixing angle. 
Finally, we investigate the new physics effects on the  polarization of the $\tau^\mp$ leptons produced in $\nu_\tau (\bar{\nu}_\tau)$ nucleon scattering.} 
\end{center}

\thispagestyle{empty} \newpage \setcounter{page}{1}
\baselineskip=14pt

\section{Introduction}

Neutrino oscillation results have confirmed that neutrinos are massive
and lepton flavors are mixed. This opens a window for searching
physics beyond the standard model (SM). Beside the standard matter effects,
the possibility of having nonstandard neutrino interactions (NSIs) is opened up.
Nonstandard neutrino interactions with matter have been extensively discussed 
\cite{Wolfenstein:1977ue,Mikheev:1986gs,Roulet:1991sm,Brooijmans:1998py, GonzalezGarcia:1998hj,Guzzo:1991hi,Bergmann:2000gp,Guzzo:2000kx,Guzzo:2001mi,Grossman:1995wx,Ota:2002na,Friedland:2005vy,Kitazawa:2006iq,Friedland:2006pi,Blennow:2007pu,EstebanPretel:2008qi,Blennow:2008ym,GonzalezGarcia:2001mp,Gago:2001xg,Huber:2001zw,Ota:2001pw,Campanelli:2002cc,Blennow:2005qj,Kopp:2007mi,Kopp:2007ne,Ribeiro:2007ud,Bandyopadhyay:2007kx,Ribeiro:2007jq,Kopp:2008ds,Malinsky:2008qn,Gago:2009ij,Palazzo:2009rb,Meloni:2009cg,Coloma:2011rq,Mitsuka:2011ty,Adhikari:2012vc,Agarwalla:2012wf,Ohlsson:2013epa}.
General bounds on NSI are summarized in Refs.~\cite{davidson, DELPHI, Biggio:2009nt}. 
The NSI impact have been studied on solar neutrinos~\cite{Berezhiani:2001rt,Friedland:2004pp,Miranda:2004nb},
atmospheric neutrinos~\cite{Bergmann:1999pk,Fornengo:2001pm,
GonzalezGarcia:2004wg,Friedland:2004ah}, reactor neutrinos \cite{Escrihuela:2009up,Barranco:2008rc}, and neutrino-nucleus scattering \cite{Barranco:2005yy, Barranco:2007tz}.

At low energy, the most general effective NSI Lagrangian
reads \cite{Kopp:2007ne}, if we consider only lepton number conserving operators, 
\begin{equation}
  \mathcal{L}_{\rm NSI} = \mathcal{L}_{V \pm A} + \mathcal{L}_{S \pm P}
                            + \mathcal{L}_T,
\end{equation}
where the different terms are classified according to their Lorentz structure
in the following way:
\begin{eqnarray}
  \mathcal{L}_{V \pm A} &=& 
    \frac{G_{F}}{\sqrt{2}} \sum_{f, f^\prime} \eps^{f,f^\prime, V \pm A}_{\alpha\beta}
      \left[ \bar{\nu}_\beta \gamma^{\rho} (1 - \gamma^{5}) \ell_\alpha \right] 
      \left[ \bar{f}^\prime \gamma_{\rho} (1 \pm \gamma^{5}) f \right]  \nonumber\\
  &+&
  \frac{G_{F}}{\sqrt{2}} \sum_{f} \! \eps^{f, V \pm A}_{\alpha\beta} \!
      \left[ \bar{\nu}_\alpha \gamma^{\rho} (1 - \gamma^{5}) \nu_\beta \right] \!\!
      \left[ \bar{f} \gamma_{\rho} (1 \pm \gamma^{5}) f \right] + {\rm h.c.},
  \label{eq:NSI-Lagrangian-VpmA} \nonumber\\
  \mathcal{L}_{S \pm P} &=&
  \frac{G_{F}}{\sqrt{2}} \sum_{f, f^\prime} \eps^{f,f^\prime, S \pm P}_{\alpha\beta}
      \left[ \bar{\nu}_\beta (1 + \gamma^{5}) \ell_\alpha \right] \!
      \left[ \bar{f}^\prime (1 \pm \gamma^{5}) f \right]+ {\rm h.c.},
  \label{eq:NSI-Lagrangian-SpmP} \nonumber\\
  \mathcal{L}_{T} &=&
  \frac{G_{F}}{\sqrt{2}} \sum_{f, f^\prime} \eps^{f,f^\prime,T}_{\alpha\beta}
      \left[ \bar{\nu}_\beta \sigma^{\rho\tau} \ell_\alpha \right] \!
      \left[ \bar{f}^\prime \sigma_{\rho\tau} f \right]+ {\rm h.c.},
  \label{eq:NSI-Lagrangian-T}
\end{eqnarray}
where  $G_F$ is the Fermi constant, $\nu_\alpha$ is the neutrino field of flavor $\alpha$, $\ell_\alpha$ is the corresponding charged lepton field, and $f$, $f^\prime$ are the components of an arbitrary weak doublet. The dimensionless NSI parameters $\eps$'s represent the strength of the nonstandard interactions relative to $G_F$  and we consider only  left-handed neutrinos. This constraint on the neutrino chirality forbids $\nu \nu f f $ terms in $\mathcal{L}_{S \pm P}$ and $\mathcal{L}_{T}$. 
If the nonstandard interactions are supposed to be mediated by a new state with a mass of order $M_{\rm NSI}$, the effective vertices in
Eq.~\eqref{eq:NSI-Lagrangian-VpmA} will be suppressed
by $1 / M_{\rm NSI}^2$ in the same way as the standard weak interactions are suppressed
by $1 / M_{\rm W}^2$. Therefore we expect that
\begin{equation}
  |\eps| \sim \frac{M_{\rm W}^2}{M_{\rm NSI}^2}.
  \label{eq:epsilon-estimate}
\end{equation}

In this work we consider the charged Higgs and $W'$ gauge boson contributions 
to neutrino-nucleon scattering. Such new states arise in many extensions of the standard model and the phenomenology of these states have been widely studied \cite{chwprime}.
In this paper we will focus on
 the $\Delta$-resonance production ($\Delta$-RES) and deep inelastic scattering (DIS) in the interactions $\scat$ and $\scatanti$ where $N=p,n$ is a nucleon and $X$ is a possible final state. In the $\Delta$-RES production we discuss the processes with $N=n,p$ and $X=\Delta^+,\Delta^0$, respectively. In the neutrino oscillation experiments, the neutrino-nucleus interaction in the detection process is assumed to be SM-like. Therefore, the extracted neutrino mixing angles, using the SM cross section, will have errors if there are new physics (NP) effects in the neutrino-nucleus amplitude. The NP effects modify the standard model cross section for $\scat$ and thus impact the extraction of the atmospheric  neutrino mixing angle $\theta_{23}$
in $\nu_{\tau}$ appearance experiments.
If high-energy Long Base Line (LBL) experiments (or atmospheric  neutrino experiments scanning in the multi-GeV neutrino energy range) could  measure $\theta_{13}$ via $\nu_{\tau}$ appearance then the NP effects in $\scat$ and $\scatanti$
would impact the $\theta_{13}$ measurement and a mismatch between this measurement and that performed at the  reactors could be a hint of a NSI in the former. The deviation of the actual mixing angle from the measured one, assuming the standard model cross section, will be studied including form factor effects in the $\Delta$-RES case.

In this paper, we make the important assumption that NP effects only arise in the coupling between the new particles and the third generation leptons, neglecting possible (subleasing) NSI effects with the first two generations. With the above assumption we can neglect NSI effects at productions since at production we have neutrino interactions involving the first and second generation leptons, only. Furthermore, the effect on $\nu_\tau$ propagation can come only from neutral current interaction. Multi Higgs models and models with $W'$ also generally contain neutral current interactions but the connection between the charged current and neutral current interactions is model dependent. In this paper we are only considering the charged current interactions, and the addition of neutral current interactions would add another model dependent parameter in our calculation. We hope to include in future work also neutral current interactions.

This pattern of NP is common in many NP models \cite{Friedland:2006pi, EstebanPretel:2008qi, Blennow:2008ym}. For instance, in multi Higgs doublet models NP effects for the third generation quarks and leptons are enhanced because of their larger masses. For the $W'$ model we are assuming a $W'$ with non-universal coupling to the generations. This is not an unusual scenario and would avoid constraints from $W'$ searches at colliders that look at the decays to $W'$ to first and second generation leptons.

%

The reaction $\scat$ is relevant to experiments like Super-Kamiokande (Super-K) \cite{Abe:2012jj, Abe:2006fu} and OPERA \cite{:2011ph} that seek to measure $\nu_{\mu} \to \nu_{\tau}$ oscillation by the observation of the $\tau$ lepton. The DONuT experiment \cite{Kodama:2007aa} measured the charged-current (CC) interaction cross section of the tau neutrino. The central-value results show  deviation from the standard model predictions by about  40\% but with large experimental errors; thus, the measurements are consistent with the standard model predictions.  
In this work we consider  NP effects within a neutrino energy range  higher than the threshold energy for the $\tau$ production where the $\Delta$-RES and DIS contributions are dominant. 
Near threshold quasielastic scattering is important.
The charged Higgs and $W'$ contributions to the quasielastic (QE) scattering $\nu_{\tau}+ n \to \tau^- + p$ and $\bar{\nu}_{\tau}+ p \to \tau^+ + n$ were considered in an earlier paper~\cite{Rashed:2012bd}. 

The hadronic transition in the charged-current (CC) interactions $\scat$ and $\scatanti$ at the partonic level is described by $(u,d) \rightarrow q$, where $q$ is a quark. In the $\Delta$-RES case $q=u,d$, while in the DIS the main contributions are obtained when $q=u,d$ because of the CKM factors. This means that the effective operator of these interactions mainly has the structure ${\cal{O}}_{NP}= \bar{u} \Gamma_i d \bar{\tau} \Gamma_j \nutau$, where $\Gamma_{i,j}$ are some Dirac structures. Therefore, we can constrain the NP parameters in this work using the constraints that have been discussed in the earlier paper~\cite{Rashed:2012bd} through the $\tau$ decay modes $\taud$ and  $\tauv$. These decay channels have operator structures similar to the one in the above CC interactions.

In Ref.~\cite{Rashed:2012bd}, we presented a model independent analysis of the NP contributions to the deviations of the mixing angles $\theta_{23}$ and $\theta_{13}$. In the case  of $\theta_{23}$, the relationship between the ratio of the NP contribution to the SM cross section $r_{23} = \sigma_{NP} (\nu_\tau) / \sigma_{SM} (\nu_\tau)$ and the deviation $\delta_{23}$ of the mixing angle was obtained in a model independent form as
\bea
\label{modineq3}
r_{23} &=&\Big[ \frac{\sin{2( \theta_{23})_{SM}}}{\sin{2( \theta_{23}) }}\Big]^2-1\,.
\eea
Here, $\theta_{23}= (\theta_{23})_{SM} +\delta_{23}$ is the actual atmospheric mixing angle, whereas $(\theta_{23})_{SM}$ is the extracted mixing angle assuming the SM $\nu_\tau$ scattering cross section and $\delta_{23}$ is the deviation . From figure (1) in Ref.~\cite{Rashed:2012bd}, one can see that  $\delta_{23}  \sim - 5^\circ$  requires $r_{23} \sim 5\%$. Similarly for $\theta_{13}$ determination, the relationship  between  $ r_{13} = \sigma_{NP}(\bar{\nu}_\tau) / \sigma_{SM}(\bar{\nu}_\tau)$ and $\delta_{13}$ is  given by
\bea
\label{modineq444}
 r_{13} &=&\Big[ \frac{\sin{2( \theta_{13})_{SM}}}{\sin{2( \theta_{13}) }}\Big]^2-1\,,
\eea
with $\theta_{13}= (\theta_{13})_{SM} +\delta_{13}$. In this case, because of the relative smallness of $\theta_{13} $ one  finds that a larger NP effect is required
to produce the deviation. As an example, $\delta_{13}  \sim - 1^\circ $  requires $ r_{13} \sim 25\%$.

A possible concern is the NP effects can be washed out after including the neutrino flux and  integrating over the possible values of the incoming neutrino energy.
We show that this is not the case by 
by considering examples of the $W'$ and charged Higgs contributions to $\delta_{23}$ using the atmospheric neutrino flux at the Super-Kamiokande experiment. The results show that the values and the pattern of the mixing angle deviation $\delta_{23}$ has no significant change due to considering the neutrino flux.

We study, also, the NP effect on the spin polarization of the produced $\tau$ lepton. The  produced $\tau$ decays to several particles  including $\nu_\tau$ and tracing back the $\tau$ decay particle distributions indicates the appearance of $\tau$. Because the $\tau$ decay distributions depend significantly on it’s spin polarization \cite{Jadach:1993hs}, the polarization information is essential to identify the $\tau$ production signal. Hence it is important to know how NP affects the $\tau$ polarization.

The paper is organized as follows: We give in the next section the kinematical relations and formalism required for $\tau$ production in the neutrino-nucleon interaction. In section (3) we present the standard model calculations for the $\Delta$-RES and DIS cross sections. In the following two sections (4, 5) we study the effects of the charged Higgs and $W'$ gauge boson contributions to the $\Delta$-RES and DIS scattering processes and the impact on the extracted neutrino mixing angles $\theta_{23}$ and $\theta_{13}$. In section (6) we study the spin polarization of the produced $\tau^\pm$ lepton. In the last section, we present our conclusions.

\section{Kinematics and formalism}

In the interactions $\scatmix$, we define the four-momenta of incoming neutrino ($k$), 
target nucleon ($p$) and produced $\tau$ lepton ($k'$) in the 
laboratory frame. The hadronic invariant mass
\bea
W^{2} =(p+q)^{2},
\eea
where $q= k- k^{\prime}$ is the four-momentum transfer,
is defined in the allowed physical region
\bea
M\leq W \leq\sqrt{s}-m_{\tau},
\label{wregion}
\eea
where $s=(k+p)^2$ is the center of mass energy and $M$ is the average nucleon mass.

The three relevant subprocesses in the neutrino-nucleon interactions are classified according to the regions of the hadronic invariant mass $W$ and the momentum transfer $q^{2}(=-Q^2)$ \cite{Hagiwara:2003di}. 
One can label QE (quasi-elastic scattering) when the hadronic invariant mass is equal to the nucleon mass $W=M$, 
RES (resonance production) when $M+m_{\pi}<W<W_{\rm cut}$, 
and IS (inelastic scattering) when  $W_{\rm cut}<W<\sqrt{s}-m_{\tau}$. 
$W_{\rm cut}$, taken in the region 1.4 GeV$\sim$1.6 GeV, is an empirical boundary between RES and IS processes, 
to avoid double counting. 
The deep inelastic scattering DIS may be labeled within the IS region when $Q^{2}\ge 1\;{\rm GeV}^{2}$, where the use of the parton model can be 
justified. 

In this paper, we consider $\Delta$-resonance state production and neglect all the other higher resonance 
states which give small contributions \cite{py,sehgal,ppy}. One can write
\beq
W^2=M^2+t+2p\cdot q,
\eeq
with $p\cdot q = M(E_\nu^{cm}-E_l^{cm})$ where the energy and momentum of the lepton and the neutrino in the center of mass (cm) system are
\bea
E_\nu^{cm} &=& \frac{(s-M^2)}{2 \sqrt{s}},\;\; p_l^{\rm cm} = \sqrt{(E_l^{cm})^2-m^2_l},\nonumber\\
E_l^{cm} &=& \frac{(s-M_\Delta^2+m^2_l)}{2 \sqrt{s}},
\eea
with $(m_l,\; M,\; M_\Delta)$ being the masses of the charged lepton, nucleon, and the $\Delta$ state, respectively. In the lab frame, the charged lepton energy is given by
\beq
E_l = \frac{t+2ME_\nu+M^2-M_\Delta^2}{2M}.
\eeq
The threshold neutrino energy to create the charged lepton partner in the $\Delta$-RES case is given by
\begin{equation}
E_{\nu_l}^{\rm th}=\frac{(m_l + M_\Delta)^2-M_n^2}{2M_n},
\end{equation}
which gives $E_{\nu_l}^{\rm th}=4.35$ GeV in the case of tau neutrino production. Using the allowed range of the invariant mass in the resonance production, the allowed region of the momentum transfer $t\equiv -Q^2$ lies in the interval
\bea
 \label{tint}
(M+m_\pi)^2 - \left(M^2+2M(E_\nu^{\rm cm}-E_l^{\rm cm}) \right)      \leq  t \leq  
W_{\rm cut}^2 - \left(M^2+2M(E_\nu^{\rm cm}-E_l^{\rm cm}) \right).
 \eea

%


\section{Standard model cross sections}

In this section we consider the standard model cross sections for the $\Delta$-RES and DIS processes. In the following sections we will study the contributions of the new states $W'$ and charged Higgs to theses two processes. In Ref.~\cite{Rashed:2012bd} we studied the NP contributions to the QE process.

\subsection{$\Delta$-Resonance production}

Neutrino-nucleon scattering produces many possible resonance states, one  of which is the $\Delta$-state. We consider here the SM cross section for the two processes which include $\nu_\tau$ and $\bar{\nu}_\tau$,
\bea
&&\scatdelta , \nonumber \\
&&\scatdeltaanti .
\eea
from the Hagiwara model \cite{Hagiwara:2003di}. That will represent the starting point of our original computation of NP effects due to charged Higgs and $W'$. Details of the SM cross section calculations can be found in Ref. \cite{Hagiwara:2003di}.
The hadronic tensor is written as
\bea
W^{\rm RES}_{\mu\nu}=\frac{\cos^{2}\theta_{c}}{4}\,
{\rm Tr}\left[P^{\beta\alpha}\Gamma_{\mu\alpha}(p\dsl+M)
\overline{\Gamma}_{\nu\beta}\right]
\frac{1}{\pi}\,\frac{W\Gamma(W)}
{(W^{2}-M_{\Delta}^{2})^{2}+W^{2}\Gamma^{2}(W)}.
\eea
Within the kinematical region of $M + m_\pi < W < W_{\rm cut}$ with $W_{\rm cut} =1.4$ GeV, we estimate the total cross section of the $\Delta$ production ($\Delta$-RES) process by integrating over $E_\tau$ and $\cos\theta$.

\subsection{Deep inelastic tau neutrino scattering}

In this section, we present the standard model cross sections for the two deep inelastic scattering (DIS) processes  which include $\nu_\tau$ and $\bar{\nu}_\tau$,
\bea
&&\scat , \nonumber \\
&&\scatanti .
\label{proc-DIS}
\eea

From Hagiwara model, see Ref. \cite{Hagiwara:2003di} for details, the differential cross section can be parametrized as follows, for $Q^2 \ll m_W^2$, 
\begin{equation}
\frac{d^2 \sigma^{{\nu}_{\tau} ({\bar{\nu}}_{\tau})}}{dx dy} = \left(\frac{G_F^2 V_{qq'}^2}{2 \pi  }\right)\;y\left(A\,W_1+\,\frac{1}{M^2}B\,W_2\,\pm \,\frac{1}{M^2}C\,W_3\,+\,\frac{1}{M^2}D\,W_5\right)\delta (\xi -x),
\label{DISSMcross}
\end{equation}
where $p_q^{\mu}=\xi p^{\mu}$ is the four-momentum of the scattering quark and $\xi$ is its momentum fraction.
The coefficients $A$,$B$,$C$,$D$ are defined as
\begin{eqnarray}
 A&=& y \left(y x + \frac{{m_l}^2}{2{E_\nu} M}\right), \nonumber\\
 B&=& \left(1-\frac{{m_l}^2}{4 {E_\nu}^2}\right) -\left(1+\frac{ M x}{2 {E_\nu}}\right)y,\nonumber\\
 C&=& 2 y \left(x\left(1-\frac{y }{2}\right)-\frac{ {m_l}^2}{4 {E_\nu} M }\right),\nonumber\\
 D&=& \frac{ {m_l}^2}{ {E_\nu} M} ,
\end{eqnarray}
where $x$ is the Bjorken variable and $y$ is the inelasticity and they are related by
\begin{equation}
 x =\frac{Q^2}{2{E_\nu} M y}.
\end{equation}
The functions $W_{1,2,3,5}$ are given in Ref. \cite{Hagiwara:2003di}.




\section{Charged Higgs contribution}

We will study the contributions of the charged Higgs to the $\Delta$-RES and DIS interactions. The deviation of the actual mixing angles $\theta_{23}$ and $\theta_{13}$, with NP contributions, from the measured ones, which assumes the SM cross section, will be discussed.

\subsection{$\Delta$-Resonance production}
\label{Charged-Higgs-RES}

We consider here the charged Higgs contribution to $\scatdelta$ and $\scatdeltaanti$. As considered in the previous paper \cite{Rashed:2012bd}, we choose the couplings of charged Higgs interactions to the SM fermions to be given by the two Higgs doublet model of type II (2HDM II) \cite{Diaz:2002tp}
\bea
\label{HiggsLag}
\mathcal{L} &=&\frac{g}{2\sqrt{2}}\left[ V_{u_i d_j} \bar{u}_i( g^{u_i d_j}_S \pm g^{u_i d_j}_P \gamma^5) d_j +  \bar{\nu}_i (g^{\nu_i l_j}_S \pm g^{\nu_i l_j}_P \gamma^5) l_j \right] \; H^{\pm}  ,
\eea
where $u_i$ and $d_j$ refer to up and down type quarks, and  $\nu_i$ and $l_j$ refer to neutrinos and the corresponding charged leptons. The other parameters are as follows: $g = e/\sin{\theta_W}$ is the SM weak coupling constant, $V_{u_i d_j}$ is the CKM matrix element, and $g_{S,P}$ are the scalar  and pseudoscalar couplings of the charged Higgs to fermions. Here, in this work, we assume the couplings $g_{S,P}$ are real and given as
\bea
\label{2HDMcoup}
g^{u_i d_j}_S &=&  \left (\frac{m_{d_j} \tan{\beta} + m_{u_i} \cot{\beta}}{M_W} \right), \nonumber\\
g^{u_i d_j}_P &=&  \left (\frac{m_{d_j} \tan{\beta} - m_{u_i} \cot{\beta}}{M_W} \right),\nonumber\\
g^{\nu_i l_j}_S &=& g^{\nu_i l_j}_P = \frac{m_{l_j} \tan{\beta}}{M_W},
\eea
where $\tan \beta$ is the ratio between the two vev's of the two Higgs doublets. From Eq.~\ref{HiggsLag} we can construct the NSI parameters  defined in Ref~\cite{Biggio:2009nt}  as $\varepsilon_{\tau\tau}^{ud(L)} \equiv \frac{m_u m_\tau}{m_H^2}$ and $\varepsilon_{\tau\tau}^{ud(R)} \equiv \frac{m_d m_\tau \tan^2 \beta}{m_H^2}$ .

The (pseudo-)scalar hadronic current $J$ for the processes 
$\scatdelta$ and $\scatdeltaanti$ in the 2HDM II is defined by 
\bea
J=\langle\Delta^{+}(p')|\hat{J}|n(p) \rangle 
=\langle\Delta^{0}(p')|\hat{J}|p(p) \rangle
=\bar{\psi}^{\alpha}_{\Delta^{+}}(p')\,\Gamma_{\alpha}\,u_{n}(p),
\eea
where the vertex $\Gamma_{\alpha}$ is expressed as
\beq
\Gamma_{\alpha} = g^{u_i d_j}_S G_V X_\alpha + g^{u_i d_j}_P G_A Y_\alpha \gamma^5 .
\eeq
Applying the equation of motion, one can obtain the hadronic matrix elements for the scalar and pseudoscalar currents
\bea
\label{vecSca}
\bra \Delta^+ (p^\prime)|\bar{u} d|n(p) \ket &=&  \bar{\psi}^\alpha_{\Delta^+} (p^\prime) G_V X_\alpha u_n (p), \nn \\
- \bra \Delta^+ (p^\prime))|\bar{u} \gamma_5 d|n(p) \ket &=&  \bar{\psi}^\alpha_{\Delta^+} (p^\prime) G_A Y_\alpha \gamma_5 u_n (p) \,,
\eea
where $X_\alpha$ and $Y_\alpha$ are 4-vectors and
\bea
G_V (t) &=&  \frac{C_5^A (t)+C_6^A (t)\; t/M^2}{m_u-m_d}, \nonumber\\
G_A (t) &=&  0,\nonumber\\
X_\alpha &=& q_\alpha .
\eea
The hadronic contribution can be written as
\bea
W^{\rm RES}=\frac{\cos^{2}\theta_{c}}{4}\,
{\rm Tr}\left[P^{\beta\alpha}\Gamma_{\alpha}(p\dsl+M)
\overline{\Gamma}_{\beta}\right]
\frac{1}{\pi}\,\frac{W\Gamma(W)}
{(W^{2}-M_{\Delta}^{2})^{2}+W^{2}\Gamma^{2}(W)}.
\eea

\begin{figure}[h!]
\centering
  \includegraphics[width=7.25cm]{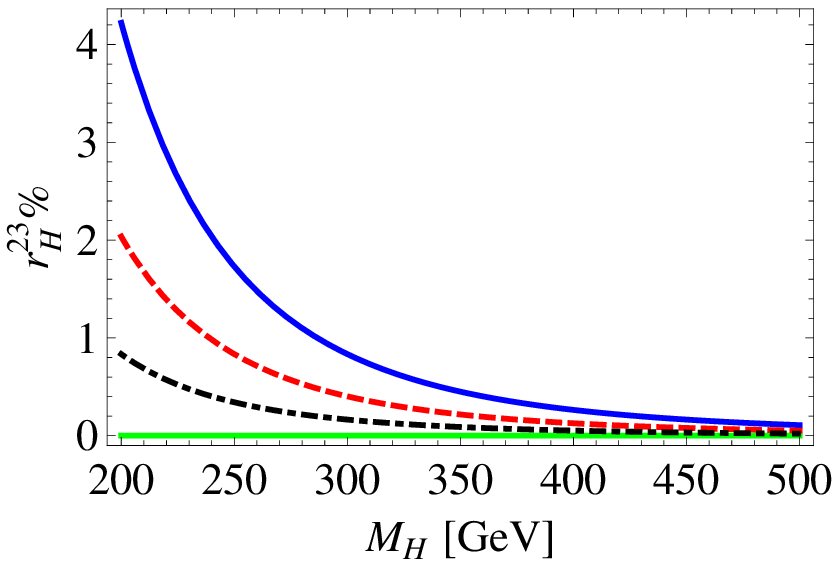}~~~
 \includegraphics[width=7.25cm]{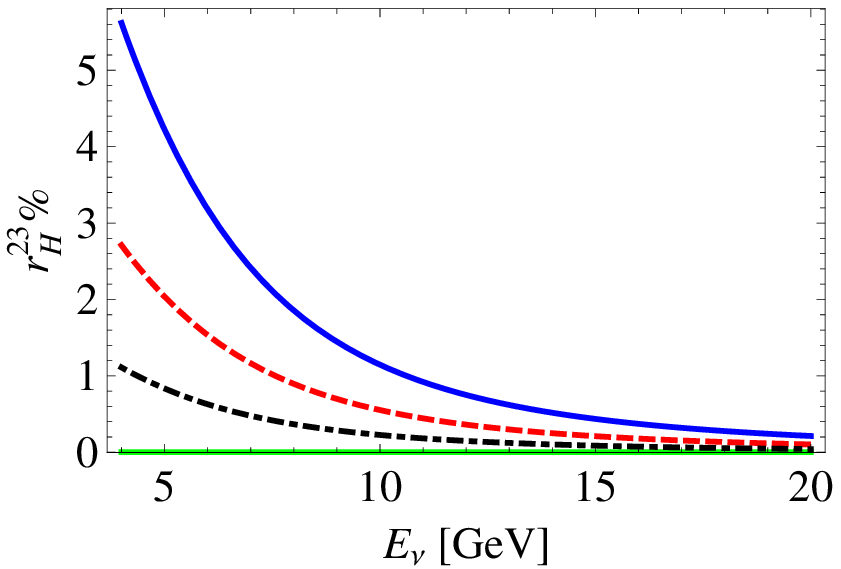}
\caption{Resonance ($H$): The figures illustrate variation of $r_H^{23} \%$ with $M_H$ (left) and $E_\nu$ (right). The green line corresponds to the SM prediction. The black (dotdashed), red (dashed), and blue (solid) lines correspond to $\tan{\beta} = 40, 50, 60$ at $E_\nu = 5$ GeV (left) and at $M_H =200$ GeV (right). }
\label{res-H-r}
\end{figure}

\begin{figure}[h!]
\centering
 \includegraphics[width=7.25cm]{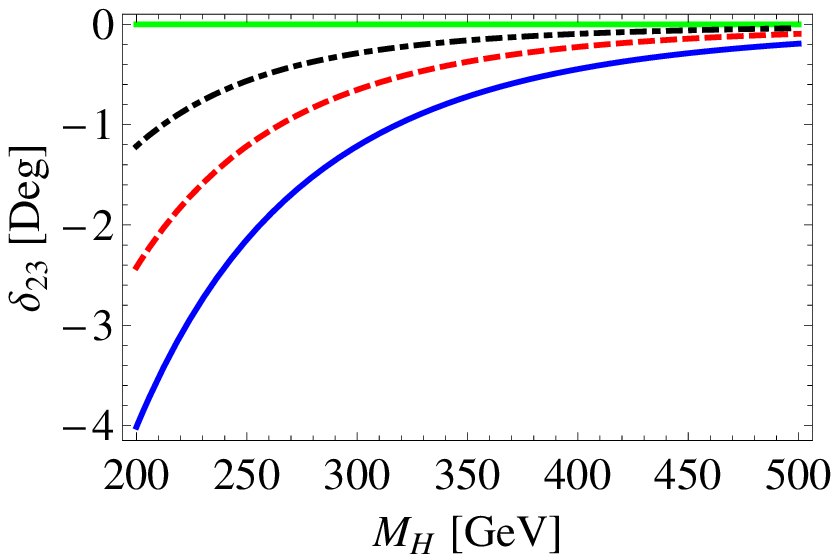}~~~
 \includegraphics[width=7.25cm]{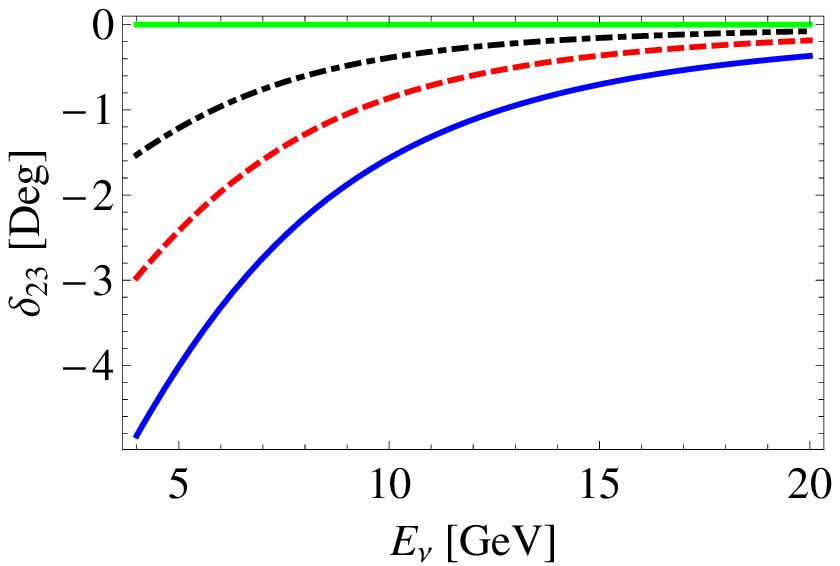}
\caption{Resonance ($H$): The figures illustrate variation of $\delta_{23}$ with $M_H$ (left) and $E_\nu$ (right). The green line corresponds to the SM prediction. The black (dotdashed), red (dashed), and blue (solid) lines correspond to $\tan{\beta} = 40, 50, 60$ at $E_\nu = 5$ GeV (left) and at $M_H =200$ GeV (right). Here, we use the best-fit value $  \theta_{23} = 42.8^\circ$ \cite{GonzalezGarcia:2010er}.}
\label{res-H-delta}
\end{figure}

\begin{figure}[h!]
\centering
 \includegraphics[width=7.25cm]{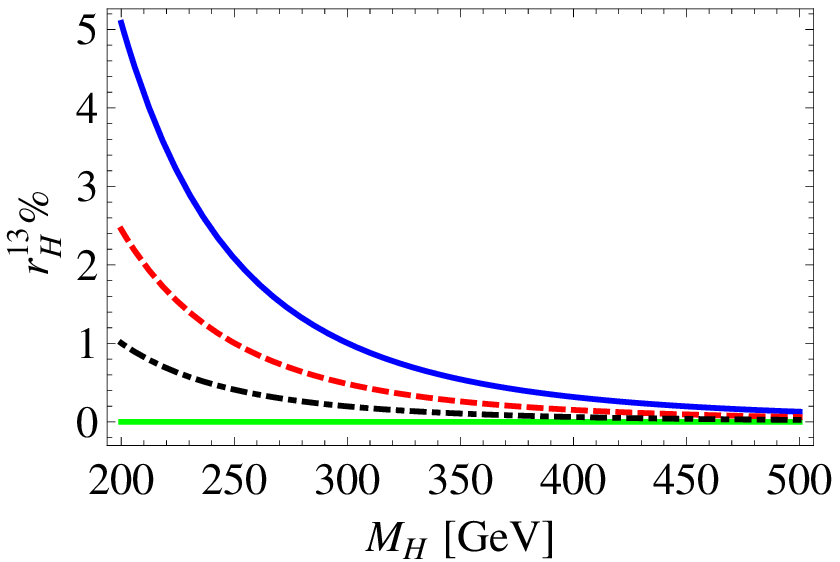}~~~
 \includegraphics[width=7.25cm]{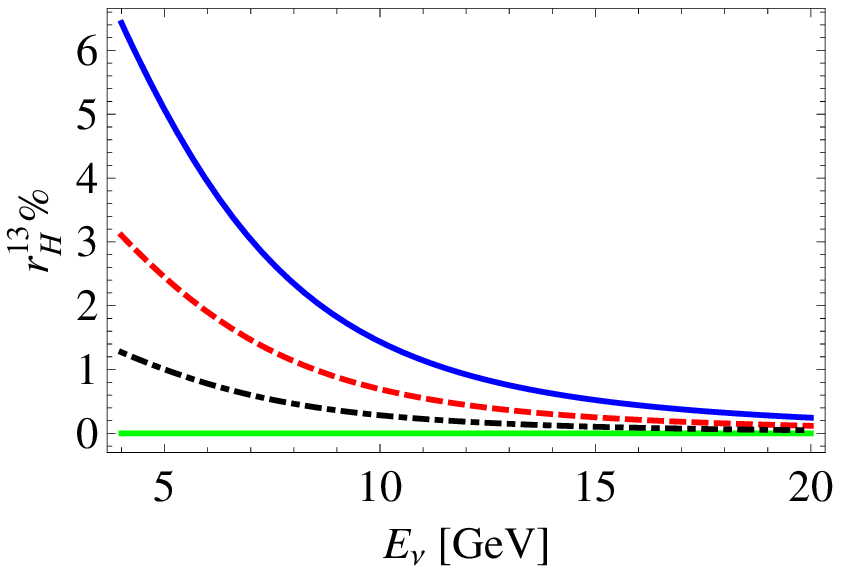}
\caption{Resonance ($H$): The figures illustrate variation of $r_H^{13} \%$ with $M_H$ (left) and $E_\nu$ (right). The green line corresponds to the SM prediction. The black (dotdashed), red (dashed), and blue (solid) lines correspond to $\tan{\beta} = 40, 50, 60$ at $E_\nu = 5$ GeV (left) and at $M_H =200$ GeV (right).}
\label{res-H-r13}
\end{figure}

\begin{figure}[h!]
\centering
 \includegraphics[width=7.25cm]{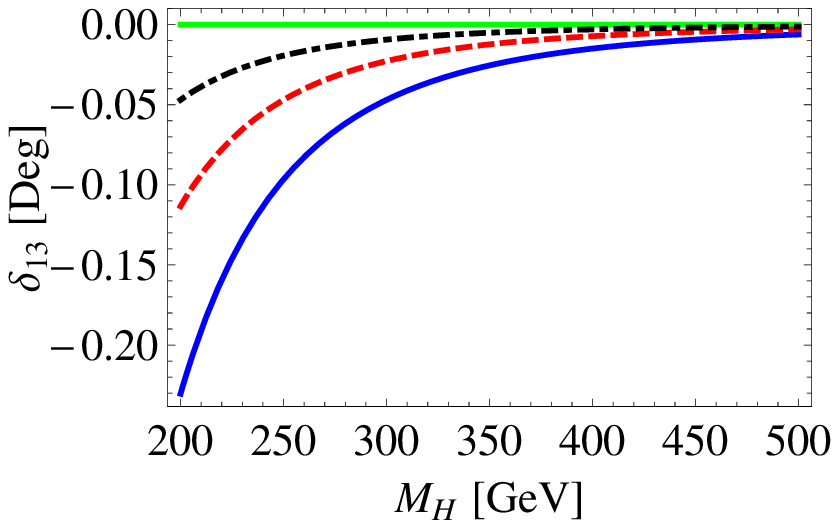}~~~
 \includegraphics[width=7.25cm]{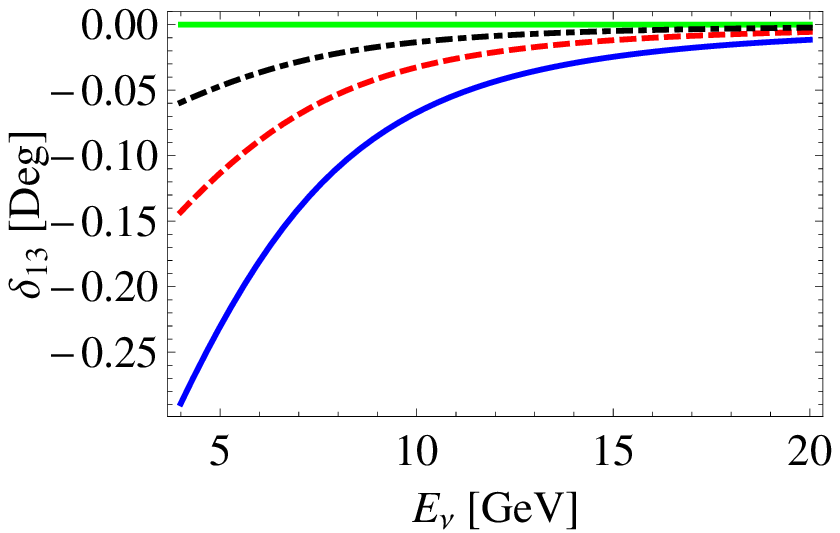}
\caption{Resonance ($H$): The figures illustrate variation of $\delta_{13}$ with $M_H$ (left) and $E_\nu$ (right). The green line corresponds to the SM prediction. The black (dotdashed), red (dashed), and blue (solid) lines correspond to $\tan{\beta} = 40, 50, 60$ at $E_\nu = 5$ GeV (left) and at $M_H =200$ GeV (right). Here, we use the best-fit value $\theta_{13} = 9.1^\circ$ \cite{Tortola:2012te}.}
\label{res-H-delta13}
\end{figure}

We use here the constraints on the NP parameters $(M_H,\; \tan \beta)$ discussed in Ref.~\cite{Rashed:2012bd} to calculate the cross sections. The ratios between the charged Higgs contributions to the two processes $\scatdelta$ and $\scatdeltaanti$ relative to the SM cross sections $r_{H}^{23} = \frac{\sigma_H (\nu_\tau)}{\sigma_{SM}(\nu_\tau)}$ and $r_{H}^{13} = \frac{\sigma_H (\bar{\nu}_\tau)}{\sigma_{SM}(\bar{\nu}_\tau)}$, respectively, can be obtained within the kinematical interval $M + m_\pi < W < 1.4$ GeV. The hadronic contribution to the matrix element is proportional to $q_\alpha(=X_\alpha)$ which varies within the small interval in Eq.~\ref{tint}. Thus, we require relatively large values of the NP parameter $\tan \beta$ to enhance the NP contributions. The ratios $r_{H}^{23}$ and $r_{H}^{13}$ decrease with increasing the incident neutrino energy and the charged Higgs mass, see Figs.~(\ref{res-H-r}, \ref{res-H-r13}). The deviations $\delta_{23}$ and $\delta_{13}$ of the atmospheric and 
reactor mixing angles, respectively, are negative as there is no interference term with the SM, see Figs.~(\ref{res-H-delta}, \ref{res-H-delta13}). Hence, the total cross sections for $\scatdelta$ and $\scatdeltaanti$ are always larger than the SM cross section. This means that, if the actual $\theta_{23}$ is close to maximal, then experiments should measure $\theta_{23}$ larger than the maximal value in the presence of a charged Higgs contribution. 
As an example, we find that $\delta_{23} \approx -5^\circ$ and $r_H^{23} \approx 6 \%$ at $E_\nu = 4$ GeV, $M_H = 200$ GeV, and $\tan \beta = 60$. 
As $\theta_{13}$ is a small angle, the deviation $\delta_{13}$ for similar set of parameters is small. For instance, we find $\delta_{13} \approx -0.3^\circ$ and $r_H^{13} \approx 6.5 \%$ at $E_\nu = 4$ GeV, $M_H = 200$ GeV, and $\tan \beta = 60$. 

\begin{figure}[h!]
\centering
 \includegraphics[width=7.25cm]{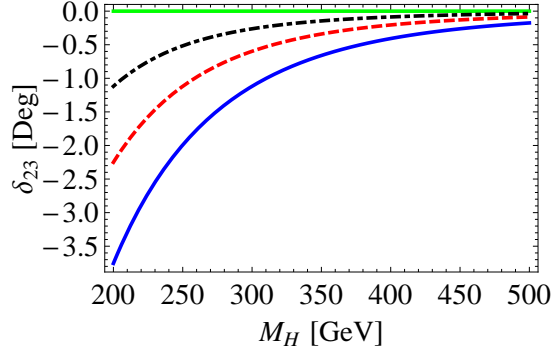}
\caption{Resonance ($H$): The figures illustrate variation of $\delta_{23}$ with $M_H$. The green line corresponds to the SM prediction. The black (dotdashed), red (dashed), and blue (solid) lines correspond to $\tan{\beta} = 40, 50, 60$. Here, we use the best-fit value $  \theta_{23} = 42.8^\circ$ \cite{GonzalezGarcia:2010er}. We take into account the atmospheric neutrino flux for Kamioka where the Super-Kamiokande experiment locates \cite{Honda:2011nf}.}
\label{Delta-RES-Flux-H}
\end{figure}

In Fig.~\ref{Delta-RES-Flux-H} we show the $\delta_{23}$ result  taking into account the atmospheric neutrino flux $\Phi (E_\nu)$ for Kamioka where the Super-Kamiokande experiment is located \cite{Honda:2011nf}. In this case the actual mixing angle $\theta_{23}$ is given as
\beq
\sin^2 2\theta_{23} = \sin^2 2\theta^{SM}_{23} R_H^{23}
\eeq
where
\beq
R_H^{23} = \int\sin^2 \frac{\Delta m_{23}^2 L}{E_\nu}  \Phi (E_\nu) \frac{d \sigma^{SM}(E_\nu, t)}{dt} dt dE_\nu / \int\sin^2 \frac{\Delta m_{23}^2 L}{E_\nu}  \Phi (E_\nu) \frac{d \sigma^{tot}(E_\nu, t)}{dt} dt dE_\nu
\eeq
with $\sigma^{tot}=\sigma^{SM}+\sigma^{NP}$. The atmospheric neutrino flux in Ref.~\cite{Honda:2011nf} is calculated averaged over all directions in the 3-dimensional scheme. We fixed the neutrino production height \cite{Honda:2004yz} at an average height with 99\% of accumulated probability for the production height. We integrate over the incoming neutrino energy from the threshold energy to 20 GeV. We find that the effect of the neutrino flux does not significantly modify the results - of order 0.1 degree.

\subsection{Deep inelastic tau neutrino scattering}
\label{Charged-Higgs-DIS}

The charged Higgs contributions to the matrix elements of the interactions $\scat$ and $\scatanti$ are given by
\begin{eqnarray}
 M_H^{\nu_\tau}&=&\left(\frac{ {G_{F}}{V_{qq'}}}{\sqrt{2}}\right)X_H\,{g^{\nu_\tau \tau}_S}\,\left[\bar{u}_\tau(k')\,(1+\gamma_5)u_{{\nu}_{\tau}}(k)\right]\;
 \left[\bar{u}_{q'}(p'_{q'})({g_{S}^{qq'}}+{g_{P}^{qq'}}\gamma_5)\;u_q(p_q)\right],\nonumber\\
 M_H^{\bar{\nu}_\tau}&=&\left(\frac{ {G_{F}}{V_{qq'}}}{\sqrt{2}}\right)X_H\,{g^{\nu_\tau \tau}_S}\,\left[\,\bar{v}_{{\nu}_{\tau}}(k)(1-\gamma_5)v_\tau(k')\right]\;
  \left[\bar{u}_{q'}(p'_{q'})({g_{S}^{qq'}}-{g_{P}^{qq'}}\gamma_5)\;u_q(p_q)\right],
\nonumber\\
\end{eqnarray}
where $q,q'=(u_i,d_j)$ and the couplings $g_{S,P}^{qq'},\; g^{\nu_\tau \tau}_S$ are defined in Eq.~\ref{2HDMcoup}.


The differential cross section is given by
\begin{eqnarray}
\frac{d^2 \sigma^{{\nu}_{\tau} ({\bar{\nu}}_{\tau})}}{dx dy} &=&\left(\frac{ G_{F}^2 V_{qq'}^2}{2\pi}\right) \; X_{H}^{2}\,({g_{S}^{v_{l}l}})^2 \;y \,L_{\mu\nu}^{{\nu}_{\tau} ({\bar{\nu}}_{\tau})} \,W^{\mu\nu} \,\delta(\xi -x) \nonumber\\
  &=&\left(\frac{ G_{F}^2 V_{qq'}^2 E_{\nu}\, M}{\pi}\right) \; X_{H}^{2}\;({g_{S}^{v_{l}l}})^2 \;\left[y \left(y x + \frac{{m_l}^2}{2{E_{\nu}} M}\right)\right]\nonumber\\
  && \frac{1}{4 }\;\left[({g_{S}^{qq'}})^2+({g_{P}^{qq'}})^2\right] \,{F_1}\,\delta(\xi -x),
\end{eqnarray}
where $X_H = M_W^2 / M_{H}^2$ and the definitions of the 2HDM coupling constants are given in Eqs.~\ref{2HDMcoup}.
There is no interference term of the SM and NP amplitudes. Thus, with the constraints on the NP parameters $(M_H,\; \tan \beta)$ \cite{Rashed:2012bd}, the charged Higgs contributions relative to the SM $r_H^{23} = \sigma_H (\nu_\tau) / \sigma_{SM} (\nu_\tau)$ and $r_H^{13} = \sigma_H (\bar{\nu}_\tau) / \sigma_{SM} (\bar{\nu}_\tau)$ are small within the kinematical interval $W_{cut} < W < \sqrt{s}-m_\tau$ GeV with $W_{cut}=1.4$ GeV.  Thus, the deviations $\delta_{23}$ and $\delta_{13}$ of the mixing angles are negligibly small.


\section{$W^\prime$ gauge boson contribution}

We study here the contributions of the $W'$ gauge boson to the $\Delta$-RES and DIS processes. The deviation of the mixing angles $\theta_{23}$ and $\theta_{13}$ will be considered.

\subsection{$\Delta$-Resonance production}
\label{Wprime-RES}

We next consider modification to the $\Delta$-RES production in $\scatdelta$ and $\scatdeltaanti$ in models with a $W^\prime$ gauge boson. The lowest dimension effective Lagrangian of $W^\prime$ interactions to the SM fermions has the form 
\bea
{\cal{L}} &=& \frac{g}{\sqrt{2}}V_{ f^\prime f} \bar{f}^\prime \gamma^\mu( g^{f^\prime f}_L P_L +  g^{f^\prime f}_R P_R) f W^\prime_\mu + ~h.c.,
\label{wprime}
\eea
where  $f^\prime$ and $f$ refer to the fermions and $g^{f^\prime f}_{L,R}$ are the left and the right handed couplings of the $W^\prime$. We will assume $g^{f^\prime f}_{L,R}$ to be real. Constraints on the couplings in Eq.~(\ref{wprime}) come from the hadronic $\tau$ decay channels $\taud$ and $\tauv$ discussed in Ref.~\cite{Rashed:2012bd}, which are consistent with the ones in Ref. \cite{Biggio:2009nt}. From Eq.~\ref{wprime}, the NSI parameters 
$\varepsilon_{\tau\tau}^{ud(L,R)}$  defined in  Ref.~\cite{Biggio:2009nt} are given as $\varepsilon_{\tau\tau}^{ud(L,R)} \equiv g_L^{\tau \nu} g_{(L,R)}^{ud}( \frac{M_W}{M_{W'}})^2 $. 

The current $J_{\mu}$ for the process 
$\scatdelta$ and $\scatdeltaanti$ in the $W'$ model is defined as
\beq
J_{\mu}=\langle\Delta^{+}(p')|\hat{J}_{\mu}|n(p) \rangle = \langle\Delta^{0}(p')|\hat{J}_{\mu}|p(p) \rangle= \bar{\psi}^{\alpha}_{\Delta^{+}}(p')\,(g_L^{ud}\Gamma_{\mu\alpha}+g_R^{ud}\Gamma^\prime_{\mu\alpha})\,u_{n}(p),
\eeq
where $\Gamma_{\mu\alpha}$ is the left-handed vertex, given in Ref.~\cite{Hagiwara:2003di}, and $\Gamma^\prime_{\mu\alpha}$ is the right-handed vertex, with ($\gamma_5 \rightarrow -\gamma_5$), for the $W'$ gauge boson. 
The hadronic tensor in the $W'$ model is now calculated from
\bea
W^{\rm RES}_{\mu\nu}&=&\frac{\cos^{2}\theta_{c}}{4}\,
{\rm Tr}\left[P^{\beta\alpha}(g_L^{ud}\Gamma_{\mu\alpha}+g_R^{ud}\Gamma^\prime_{\mu\alpha})(p\dsl+M)
(g_L^{ud}\bar{\Gamma}_{\nu\beta}+g_R^{ud}\bar{\Gamma}^\prime_{\nu\beta})\right]\nonumber\\
&&\frac{1}{\pi}\frac{W\Gamma(W)}
{(W^{2}-M_{\Delta}^{2})^{2}+W^{2}\Gamma^{2}(W)}.
\eea

\begin{figure}[h!]
\centering
 \includegraphics[width=7.25cm]{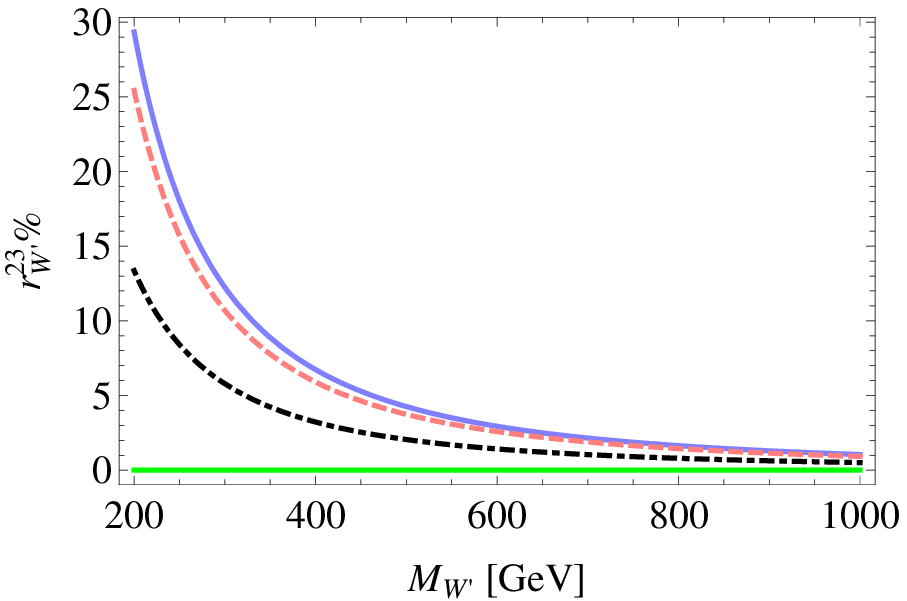}~~~
 \includegraphics[width=7.25cm]{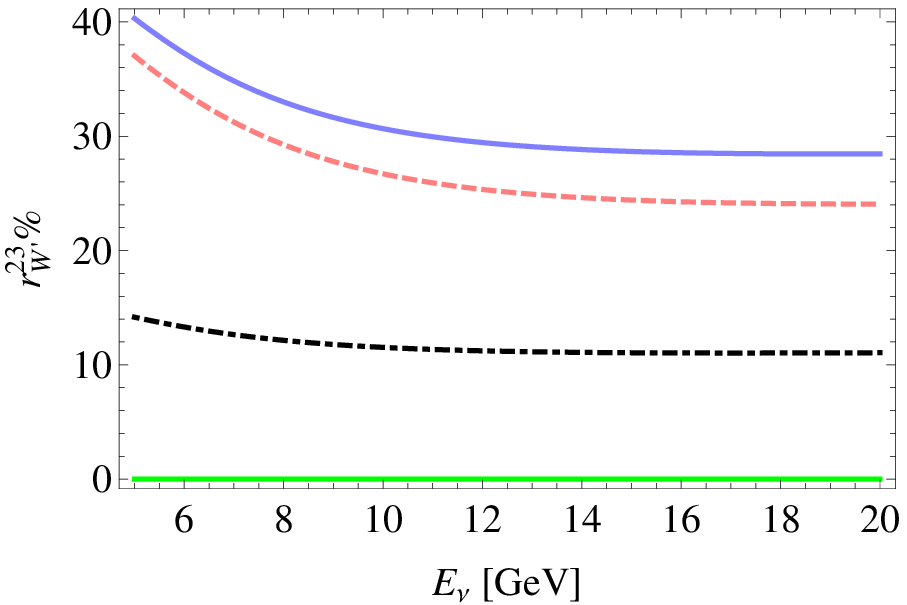}
\caption{Resonance ($W^\prime$): The left (right) panel figures illustrate the variation of $r_{W'}^{23}\%$ with the $W^\prime$ mass $M_{W'}$ ($E_\nu$) when both left and right-handed  $W^\prime$ couplings are present. The lines show predictions for some representative values of the $W^\prime$ couplings $(g^{\tau \nu_\tau}_L, g^{ud}_L, g^{ud}_R)$. The green line (solid, lower) corresponds to the SM prediction. The blue line (solid, upper) in the left  figure corresponds to (-0.94 ,  -1.13 , -0.85)  at $E_{\nu} = 17$  GeV, and the blue line (solid, upper) in the right  figure corresponds to (1.23 , 0.84 , 0.61) at $ M_{W^\prime} = 200 $ GeV.}
\label{r-Wp-Res}
\end{figure}

\begin{figure}[h!]
\centering
 \includegraphics[width=7cm]{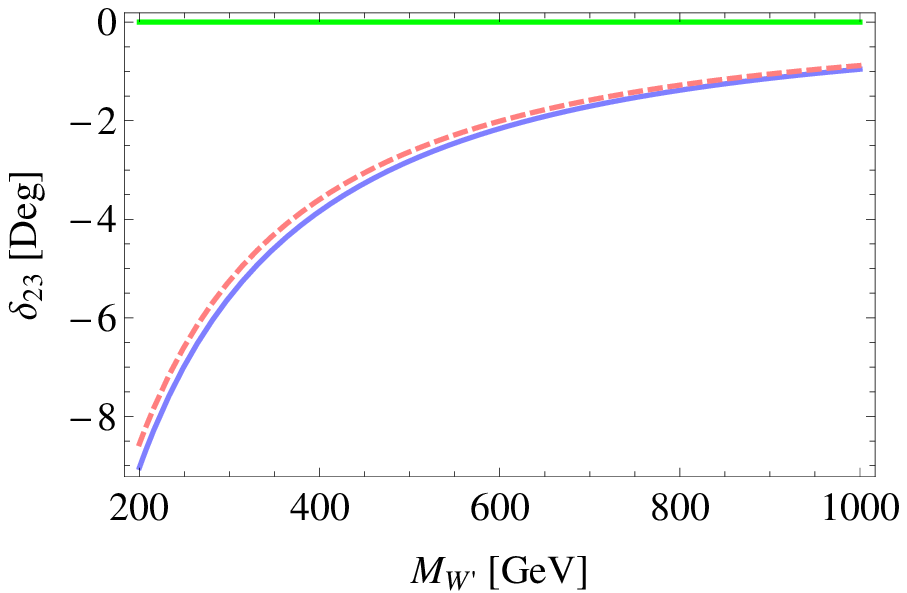}~~
 \includegraphics[width=7cm]{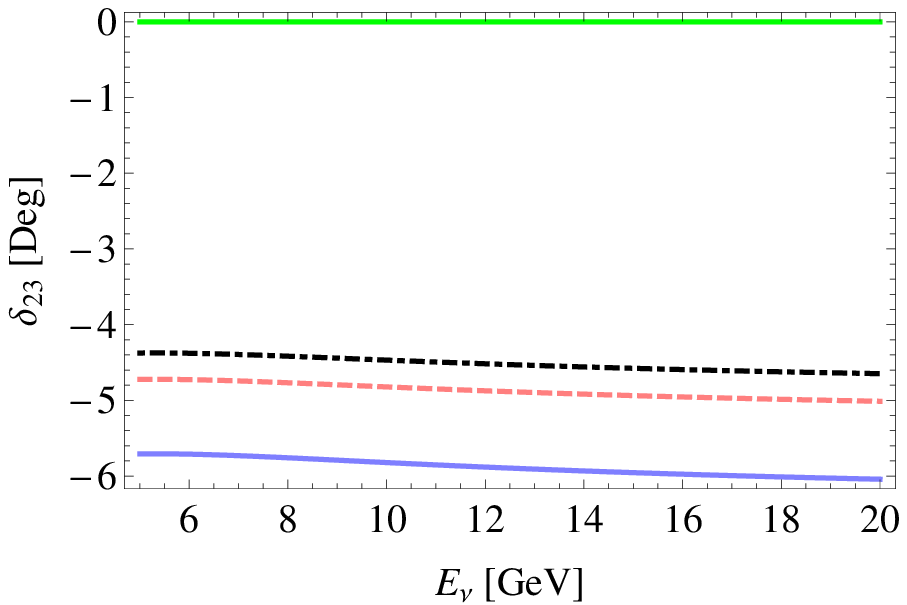}
\caption{Resonance ($W^\prime$): The left (right) panel figures illustrate the deviation $\delta_{23}$ with the $W^\prime$ mass $M_{W'}$ ($E_\nu$) when only left-handed $W^\prime$ couplings are present. The lines show predictions for some representative values of the $W^\prime$ couplings $(g^{\tau \nu_\tau}_L, g^{ud}_L)$. The green line (solid, upper) corresponds to the SM prediction. The blue line (solid, lower) in the left  figure corresponds to (0.69, 0.89)  at $E_{\nu} = 17$  GeV, and the blue line (solid, lower) in the right figure corresponds to (1.42, 0.22) at $ M_{W^\prime} = 200 $ GeV. Here, we use the best-fit value $  \theta_{23} = 42.8^\circ$ \cite{GonzalezGarcia:2010er}.}
\label{delta-Wp-Res-RH}
\end{figure}

\begin{figure}[h!]
\centering
 \includegraphics[width=7cm]{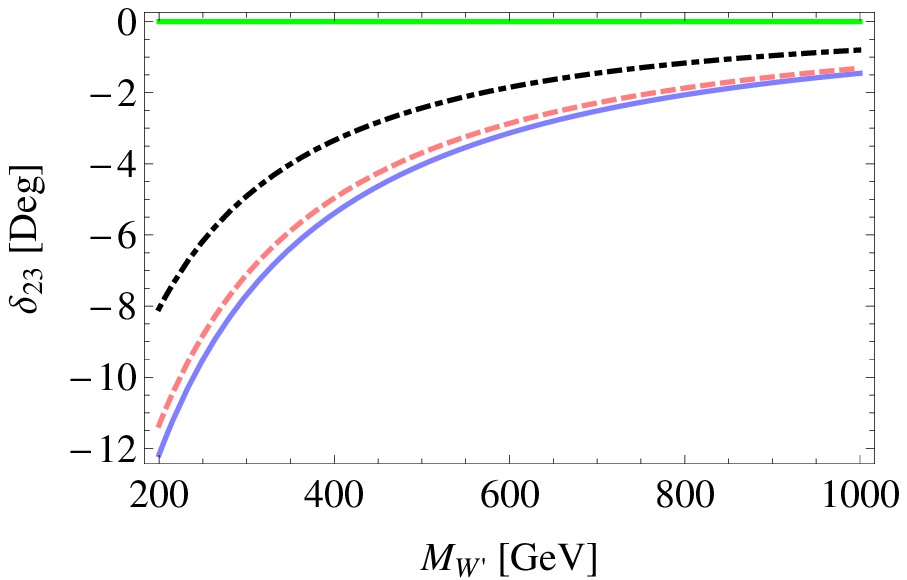}~~
 \includegraphics[width=7cm]{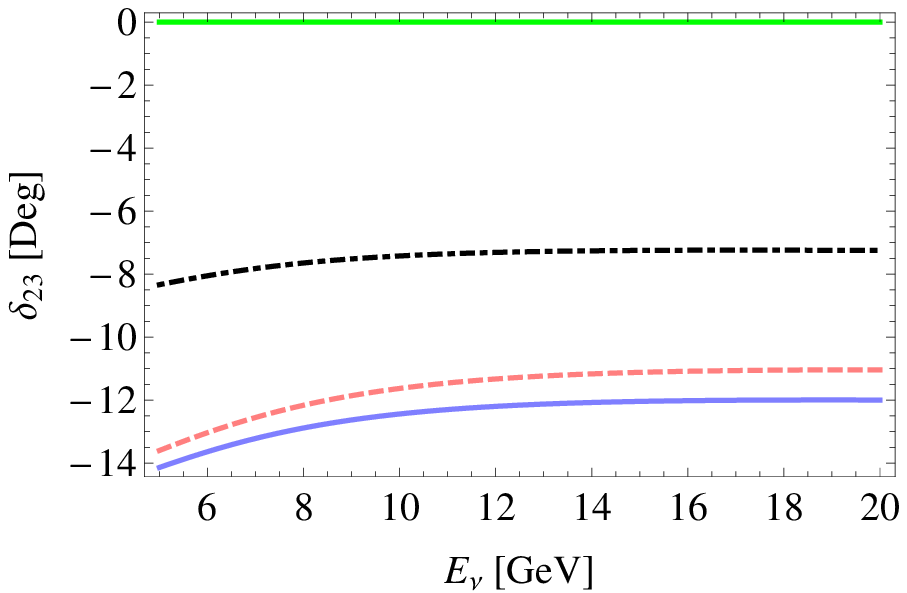}
\caption{Resonance ($W^\prime$): The left (right) panel figures illustrate the deviation $\delta_{23}$ with the $W^\prime$ mass $M_{W'}$ ($E_\nu$) when both left and right-handed  $W^\prime$ couplings are present. The lines show predictions for some representative values of the $W^\prime$ couplings $(g^{\tau \nu_\tau}_L, g^{ud}_L, g^{ud}_R)$. The green line (solid, upper) corresponds to the SM prediction. The blue line (solid, lower) in the left  figure corresponds to (-0.94 ,  -1.13 , -0.85)  at $E_{\nu} = 17$  GeV, and the blue line (solid, lower) in the right  figure corresponds to (1.23 , 0.84 , 0.61) at $ M_{W^\prime} = 200 $ GeV. Here, we use the best-fit value $  \theta_{23} = 42.8^\circ$ \cite{GonzalezGarcia:2010er}.}
\label{delta-Wp-Res}
\end{figure}

\begin{figure}[h!]
\centering
 \includegraphics[width=7.25cm]{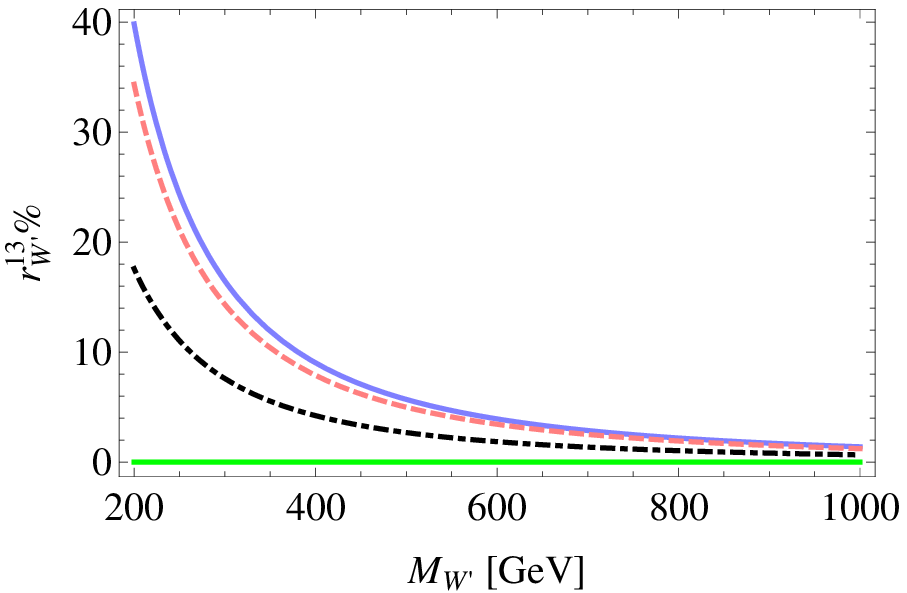}~~~
 \includegraphics[width=7.25cm]{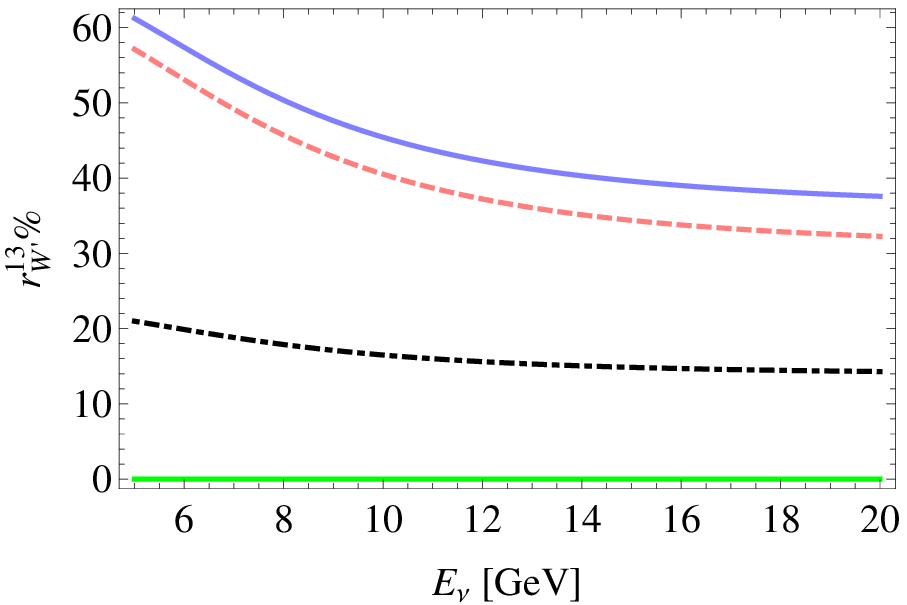}
\caption{Resonance ($W'$): The left (right) panel figures illustrate the variation of $r_{W'}^{13}\%$ with the $W^\prime$ mass $M_{W'}$ ($E_\nu$) when both left and right-handed  $W^\prime$ couplings are present. The lines show predictions for some representative values of the $W^\prime$ couplings $(g^{\tau \nu_\tau}_L, g^{ud}_L, g^{ud}_R)$. The green line (solid, lower) corresponds to the SM prediction. The blue line (solid, upper) in the left  figure corresponds to (-0.94 ,  -1.13 , -0.85)  at $E_{\nu} = 17$  GeV, and the blue line (solid, upper) in the right  figure corresponds to (1.23 , 0.84 , 0.61) at $ M_{W^\prime} = 200 $ GeV.}
\label{res-W-r13}
\end{figure}

\begin{figure}[h!]
\centering
 \includegraphics[width=7.25cm]{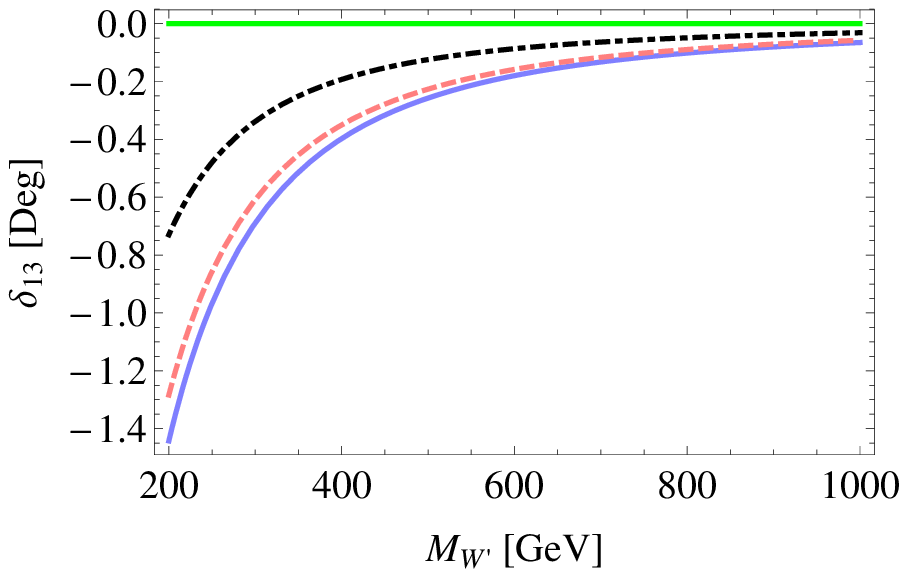}~~~
 \includegraphics[width=7.25cm]{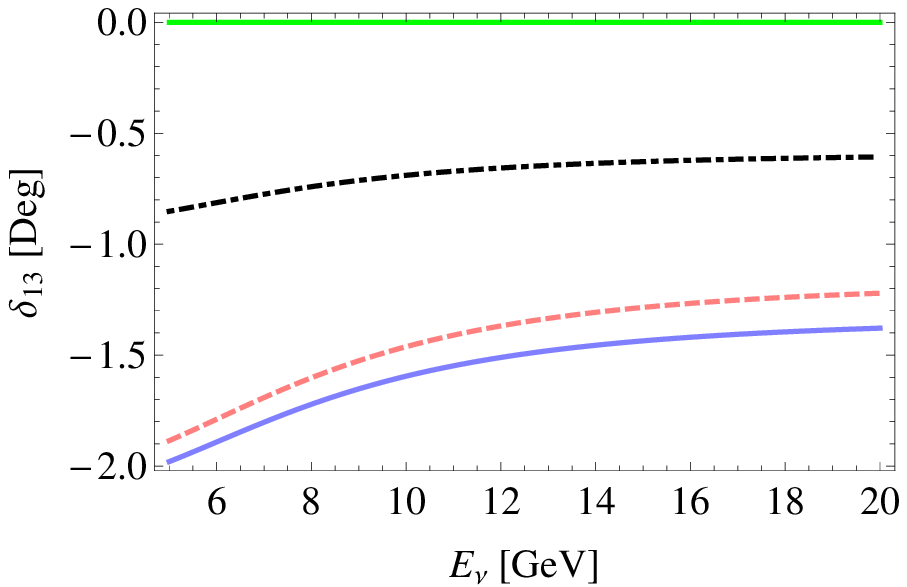}
\caption{Resonance ($W'$): The left (right) panel figures illustrate the deviation $\delta_{13}$ with the $W^\prime$ mass $M_{W'}$ ($E_\nu$) when both left and right-handed  $W^\prime$ couplings are present. The lines show predictions for some representative values of the $W^\prime$ couplings $(g^{\tau \nu_\tau}_L, g^{ud}_L, g^{ud}_R)$. The green line (solid, upper) corresponds to the SM prediction. The blue line (solid, lower) in the left  figure corresponds to (-0.94 ,  -1.13 , -0.85) at $E_{\nu} = 17$  GeV, and the blue line (solid, lower) in the right  figure corresponds to (1.23 , 0.84 , 0.61) at $ M_{W^\prime} = 200 $ GeV. Here, we use the best-fit value $  \theta_{13} = 9.1^\circ$ \cite{Tortola:2012te}. }
\label{res-W-delta13}
\end{figure}

\begin{figure}[h!]
\centering
 \includegraphics[width=7cm]{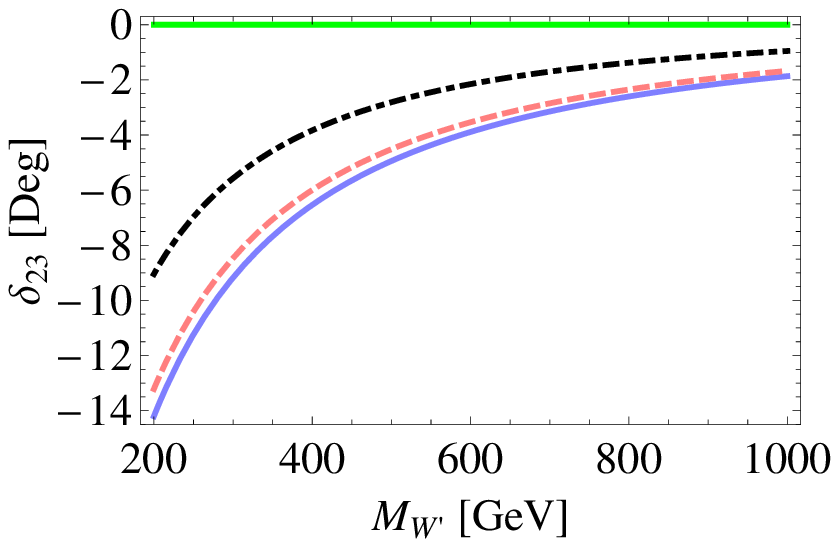}
\caption{Resonance ($W^\prime$): The figure illustrates the deviation $\delta_{23}$ with the $W^\prime$ mass $M_{W'}$ when both left and right-handed  $W^\prime$ couplings are present. The lines show predictions for some representative values of the $W^\prime$ couplings $(g^{\tau \nu_\tau}_L, g^{ud}_L, g^{ud}_R)$. The green line (solid, upper) corresponds to the SM prediction. The blue line (solid, lower) corresponds to (-0.94 ,  -1.13 , -0.85). Here, we use the best-fit value $  \theta_{23} = 42.8^\circ$ \cite{GonzalezGarcia:2010er}. We take into account the atmospheric neutrino flux for Kamioka where the Super-Kamiokande experiment locates \cite{Honda:2011nf}.}
\label{Delta-RES-Flux-Wp}
\end{figure}

Using the constraints on the $W'$ couplings discussed in Ref.~\cite{Rashed:2012bd}, the ratios of the $W'$ contributions to $\scatdelta$ and $\scatdeltaanti$ relative to the SM cross sections $r_{W'}^{23} = \frac{\sigma_{W'} (\nu_\tau)}{\sigma_{SM}(\nu_\tau)}$ and $r_{W'}^{13} = \frac{\sigma_{W'} (\bar{\nu}_\tau)}{\sigma_{SM}(\bar{\nu}_\tau)}$, respectively, are shown in Figs.~(\ref{r-Wp-Res}, \ref{res-W-r13}). The $r_{W'}^{23}$ and $r_{W'}^{13}$ values are mostly positive which, in turn, leads to $\delta_{23}$ and $\delta_{13}$ being mostly negative. The variation of $\delta_{23}$ and $\delta_{13}$ with the $W'$ mass and $E_\nu$ in the SM-like case, with only left-handed couplings, and for the case where both the LH and RH couplings are present are shown in Figs.~(\ref{delta-Wp-Res-RH}, \ref{delta-Wp-Res}, \ref{res-W-delta13}). 
As a typical example, we find that $\delta_{23} \approx -14^\circ$ at $E_\nu = 4$ GeV, $M_{W'} = 200$ GeV, and $(g^{\tau \nu_\tau}_L, g^{ud}_L, g^{ud}_R)=(1.23,0.84,0.61)$. 
Because of the smallness of $\theta_{13}$, the NP effect on the extraction of $\theta_{13}$ is small. Achieving large $\delta_{13}$ within the constraints given in Ref.~\cite{Rashed:2012bd} is difficult in this model.
As an example, we find that $\delta_{13} \approx -2^\circ$ at $E_\nu = 4$ GeV, $M_{W'} = 200$ GeV, and $(g^{\tau \nu_\tau}_L, g^{ud}_L, g^{ud}_R)=(1.23,0.84,0.61)$. In Fig.~\ref{Delta-RES-Flux-Wp}, the results show small modification to the $\delta_{23}$ values when considering the atmospheric neutrino flux \cite{Honda:2011nf} - of the size of one degree.

\subsection{Deep inelastic tau neutrino scattering}
\label{Wprime-DIS}

The matrix elements are
\begin{eqnarray}
 M_{W'}^{\nu_\tau}&=&\left(\frac{-i {G_{F}}{V_{qq'}}K_{W'}}{\sqrt{2}}\right)\;\left[\bar{u}_\tau(k')\gamma^\mu(1-\gamma_5)u_{{\nu}_{\tau}}(k)\right]\;\left[ \bar{u}_{q'}(p'_{q'})\;{\gamma}_\mu\left({\gamma}_{W'}^\rho-
 {\gamma}_{W'}^\kappa\gamma_5\right)u_q(p_q)\right] ,\nonumber\\
 M_{W'}^{\bar{\nu}_\tau}&=&\left(\frac{-i {G_{F}}{V_{qq'}}K_{W'}}{\sqrt{2}}\right)\;\left[\bar{v}_{{\nu}_{\tau}}(k)\gamma^\mu(1-\gamma_5) v_\tau (k')\right]\;\left[ \bar{u}_{q'}(p'_{q'})\;{\gamma}_\mu\left({\gamma}_{W'}^\rho-
  {\gamma}_{W'}^\kappa\gamma_5\right)u_q(p_q)\right] ,\nonumber\\
\end{eqnarray}
where the definitions are 
\begin{eqnarray}
 {\gamma}_{W'}^\rho &=& X_{W'}g_L^{{\nu}_\tau \tau}(g_L^{qq'}+g_R^{qq'}), \nonumber\\
 {\gamma}_{W'}^\kappa &=& X_{W'}g_L^{\nu_\tau \tau}(g_L^{qq'}-g_R^{qq'}), \nonumber\\
 X_{W'}&=& \left(\frac{m_{W}^2}{m_{W'}^2}\right),\nonumber\\
 K_{W'} &=& \left(1+\frac{Q^2}{m_{W'}^2}\right)^{-1}.
\end{eqnarray}
The total differential cross section has the same form as the SM one in Eq.~(\ref{DISSMcross}), after setting $K_{W'}^2\sim 1$, 
\begin{equation}
\label{DISSMandWprimecross}
\frac{d^2 \sigma^{{\nu}_{\tau} ({\bar{\nu}}_{\tau})}_{SM+W^\prime}}{dx dy} = \left(\frac{G_F^2 V_{qq'}^2}{2 \pi  }\right)\;y\left(A^\prime \,W_1+\,\frac{1}{M^2}B^\prime\,W_2\,\pm \,\frac{1}{M^2}C^\prime \,W_3\,+\,\frac{1}{M^2}D^\prime \,W_5\right)\delta (\xi -x),
\end{equation}
where $A^\prime$,$B^\prime$,$C^\prime$, and $D^\prime$ are defined as: 
\begin{eqnarray}
 A'&=& \frac{1}{2} A \left(|a'|^2+|b'|^2 \right), \nonumber\\
 B'&=& \frac{1}{2} B \left(|a'|^2+|b'|^2\right) , \nonumber\\
 C'&=& Re[a' b'^*]C , \nonumber\\
 D'&=&\frac{1}{2}  D \left(|a'|^2+|b'|^2 \right).
 \label{primeprime}
\end{eqnarray}
with 
\bea
a' &=& 1 + {\gamma}_{W'}^\rho,\nonumber\\
b' &=& 1 + {\gamma}_{W'}^\kappa.
\eea

The ratios of the $W'$ contributions to the SM cross sections $r_{W'}^{23}$ and $r_{W'}^{13}$ and the deviations $\delta_{23}$ and $\delta_{13}$ are shown within the allowed kinematical range $M + m_\pi < W < 1.4$ GeV in  Figs.~(\ref{res-W-r23-new}, \ref{res-W-delta23-new}, \ref{res-W-r13-new}, \ref{res-W-delta13-new}). The $r_{W'}^{23}$ and $r_{W'}^{13}$ values are mostly positive which, in turn, leads to $\delta_{23}$ and $\delta_{13}$ being mostly negative, respectively. 
As some examples,
we find that $\delta_{23} \approx -14^\circ$ and $\delta_{13} \approx -1.5^\circ$ at $E_\nu = 17$ GeV, $M_{W'} = 200$ GeV, and $(g^{\tau \nu_\tau}_L, g^{ud}_L, g^{ud}_R)=(-0.94,-1.13,-0.85)$. 
In Fig.~\ref{DIS-Flux-Wp}, the results show a negligible change to the $\delta_{23}$ values when considering the atmospheric neutrino flux \cite{Honda:2011nf}.

Finally, we note that one could detect the presence of NSI's by comparing the number of observed events to the number of expected events based on the SM.
One can calculate the number of events in the SM  as $N_{\rm SM}\pm \Delta N_{\rm SM}$ where $\Delta N_{\rm SM}$  is the error in the SM estimation of the number of events. If the number of events estimated in the $W'$ model $N_{\rm NSI}$ falls beyond the uncertainty of the SM measurement, then  the impact of NSI is large
enough to be detectable at neutrino oscillation experiments.\\
The rate of change of the observed electron and muon-neutrino scattering cross sections with respect to the neutrino energy  become constant at high energies \cite{PDG-No}, i.e. $\sigma_{\nu_{e/\mu}}(E)=\left(d\sigma_{\nu_{e,\mu}}/dE\right) ^{\rm const} E$. Because of the kinematic effects due to the $\tau$-lepton mass, the $\nu_\tau$ cross section can be parametrized as \cite{Kodama:2007aa}
\beq
\sigma_{\nu_\tau}^{\rm SM}=\left(\frac{d\sigma_{\nu_\tau}}{dE}\right)^{\rm const} E K(E),
\eeq
where $\left(d\sigma_{\nu_\tau}/dE\right)^{\rm const}$ is the energy-independent factor of the cross section, and $K$ gives the part of the tau-neutrino cross section that depends on kinematic effects due to the $\tau$-lepton mass. From the measured muon-neutrino cross section in the PDG \cite{PDG-No}, $\left(d\sigma_{\nu_\mu}/dE\right)^{\rm const}=(0.51\pm 0.056)\times 10^{-38}$ cm$^2$/GeV. The average  error ($0.056\times 10^{-38}$ cm$^2$/GeV) of the cross-section  includes the systematic, statistical, and normalization uncertainties and has been taken for neutrino energies above 30 GeV, where the DIS contribution is dominant. For instance, the measured muon-neutrino scattering cross section at MINOS experiment \cite{Adamson:2009ju} provides an explicit value $\left(d\sigma_{\nu_\mu}/dE\right)^{\rm const}=(0.675\pm 0.012\pm 0.004\pm 0.011)\times 10^{-38}$ cm$^2$/GeV with the uncertainty types statistical, systematic, and normalization resulting in the total uncertainty 0.018 for the energy range 30-50 GeV (the MINOS results are included in the average value).  Since we consider the NP contributions in the tau sector only, we can 
take  the energy-independent factor of the SM tau-neutrino cross section to be given as $\left(d\sigma_{\nu_\tau}/dE\right)^{\rm const}=\left(d\sigma_{\nu_\mu}/dE\right)^{\rm const}$ because of the SM universality of the weak interactions. The uncertainty of the number of $\nu_\tau$ events calculated in the SM limit follow from the uncertainty of $\left(d\sigma_{\nu_\tau}/dE\right)^{\rm const}$.

The number of tau-neutrino events in the SM is found to be $N_{\rm SM}=30.66\pm 3.37$ using the PDG cross section value  for the 22.5 kton fiducial volume of the Super-K detector \cite{Abe:2012jj} during the 2806 day running period. 
%
The atmospheric neutrino flux \cite{Honda:2004yz} has been taken  for vertically upward going neutrinos ($\cos \theta =-1$) where $\theta$ is the zenith angle. The distance $d$ traveled by atmospheric neutrinos can be calculated by  \cite{Honda:2004yz}
\beq
d=\sqrt{(h^2+2R_e h)+(R_e \cos\theta)^2}-R_e \cos\theta,
\eeq 
where $h$ is the neutrino production height and $R_e$ is the radius of the earth - its surface is assumed to be spherical. In Ref.~\cite{Honda:2004yz} there is a distribution for the atmospheric neutrino flux at a zenith angle around ($\cos \theta=1$). Since the distribution of the flux over the zenith angle is symmetric at high neutrino energy, see \cite{Conrad:2010mh}, the flux is the same at $\cos \theta=-1$ and $1$. We choose to work with $\cos \theta=-1$ because the distance $d$ will be maximum, through the diameter of the earth, which in turns enhances the transition probability
We take $h=4.5\times 10^4$ m  for the accumulated probability of $99\%$ for the vertical production height \cite{Honda:2004yz} and we integrate over the neutrino energies from $30-100$ GeV.

\begin{figure}[h!]
\centering
 \includegraphics[width=7cm]{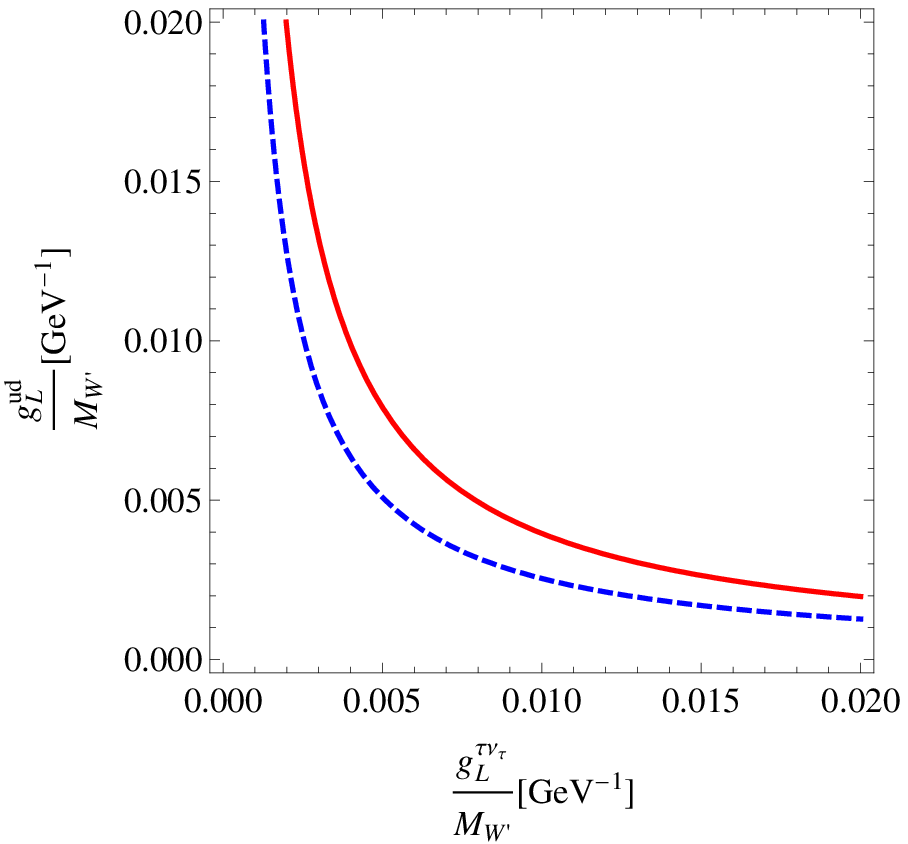}~~
 \includegraphics[width=7cm]{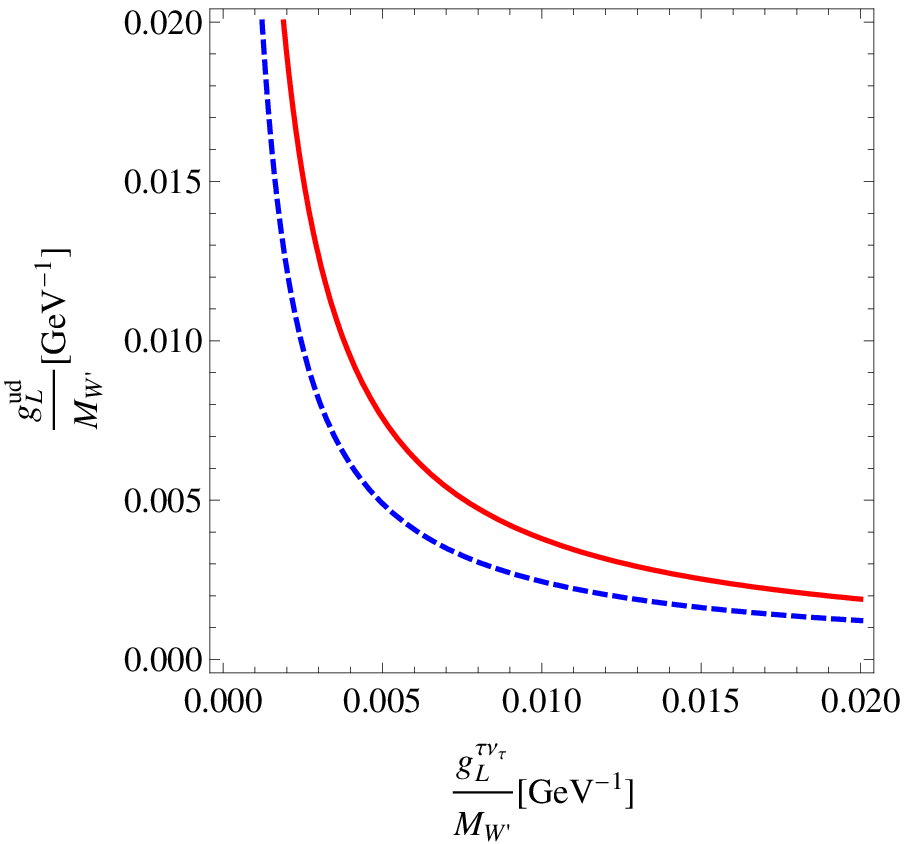}
\caption{Contour plot for $3\sigma$ (blue dashed) and $5\sigma$ (red solid) for the number of events in the presence of NSI. The left panel is for $g_R^{ud}=0$ and the right panel is for $g_R^{ud}=g_L^{ud}$.}
\label{contour}
\end{figure}

Next we calculate the number of events in the $W'$ model in the DIS energy region ignoring  the QE and $\Delta$-RES contributions. In order to cross-check our calculations we find the SM number of events to be $N_{SM}=30.08$ by setting the couplings $(g_L^{\nu_\tau \tau}, g_L^{ud}, g_R^{ud})=(0,0,0)$ which is very close to $N_{\rm SM}$ estimated above. 
In Fig.~\ref{contour}, we show the contour plot for the number of events in the presence of the NSI. We use the $\chi^2$ measure to make the $3\sigma$ and $5\sigma$ plots where
\beq
\chi^2 (M_{W'},g_L^{\nu_\tau \tau},g_L^{ud},g_R^{ud})=\frac{\left[ N_{NSI}(M_{W'},g_L^{\nu_\tau \tau},g_L^{ud},g_R^{ud})-N_{SM}\right]^2}{\sigma^2},
\eeq
and $\sigma=3.37$ is the standard deviation. We calculate the contour plots for $N_{NSI}=40.75$ and $N_{NSI}=47.48$ which are 3$\sigma$ and $5\sigma$, respectively, away from the SM prediction $N_{\rm SM}=30.66\pm 3.37$. In Fig.~\ref{contour} left panel we assume non-zero value for the left handed coupling and a vanishing right handed coupling $g_R^{ud}=0$, while in the right panel we assume $g_R^{ud}=g_L^{ud}$. In the DIS cross section, we find that the cross section is symmetric under the interchange of $a'$ and $b'$, see Eq.~\ref{primeprime}. This means that the contour plot for $g_R^{ud}=g_L^{ud}$ and $g_R^{ud}=-g_L^{ud}$ are the same.

\begin{figure}[h!]
\centering
 \includegraphics[width=7.25cm]{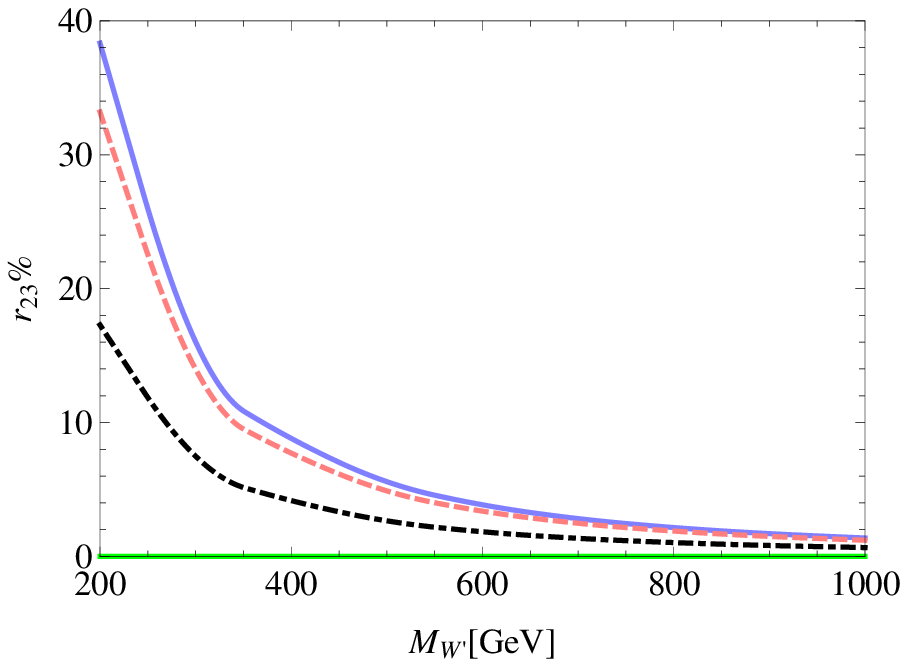}~~~
 \includegraphics[width=7.25cm]{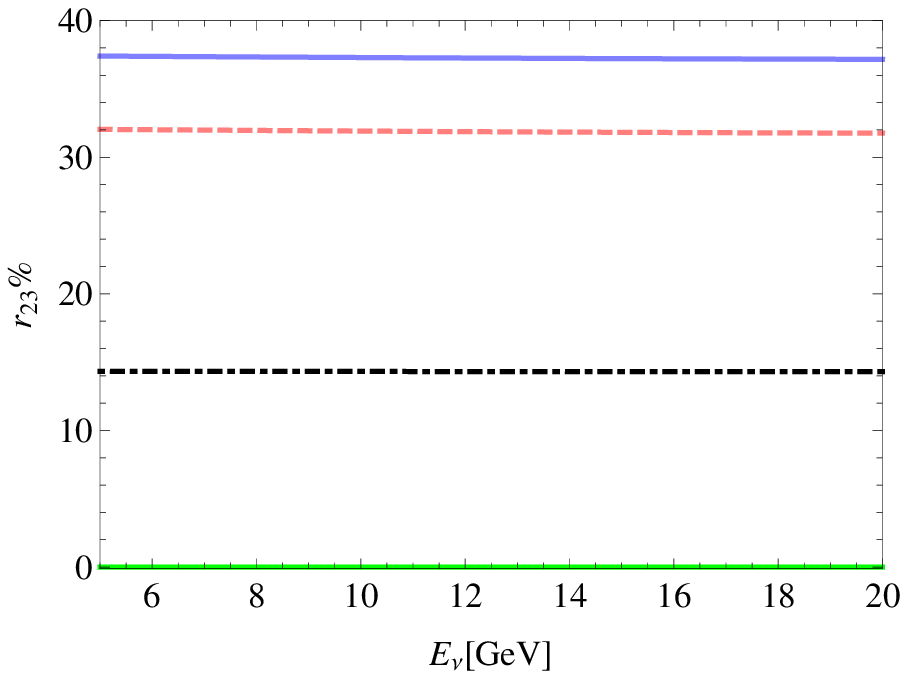}
\caption{DIS ($W'$): The left (right) panel figures illustrate the variation of $r_{W'}^{23}\%$ with the $W^\prime$ mass $M_{W'}$ ($E_\nu$) when both left and right-handed  $W^\prime$ couplings are present. The lines show predictions for some representative values of the $W^\prime$ couplings $(g^{\tau \nu_\tau}_L, g^{ud}_L, g^{ud}_R)$. The green line (solid, lower) corresponds to the SM prediction. The blue line (solid, upper) in the left  figure corresponds to (-0.94 ,  -1.13 , -0.85)  at $E_{\nu} = 17$  GeV, and the blue line (solid, upper) in the right  figure corresponds to (1.23 , 0.84 , 0.61) at $ M_{W^\prime} = 200 $ GeV.}
\label{res-W-r23-new}
\end{figure}

\begin{figure}[h!]
\centering
 \includegraphics[width=7.25cm]{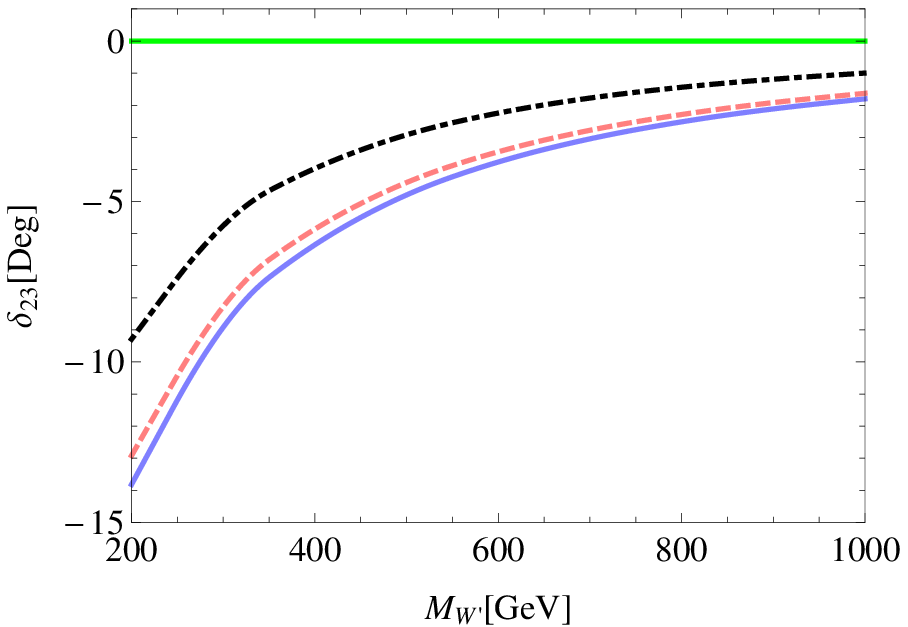}~~~
 \includegraphics[width=7.25cm]{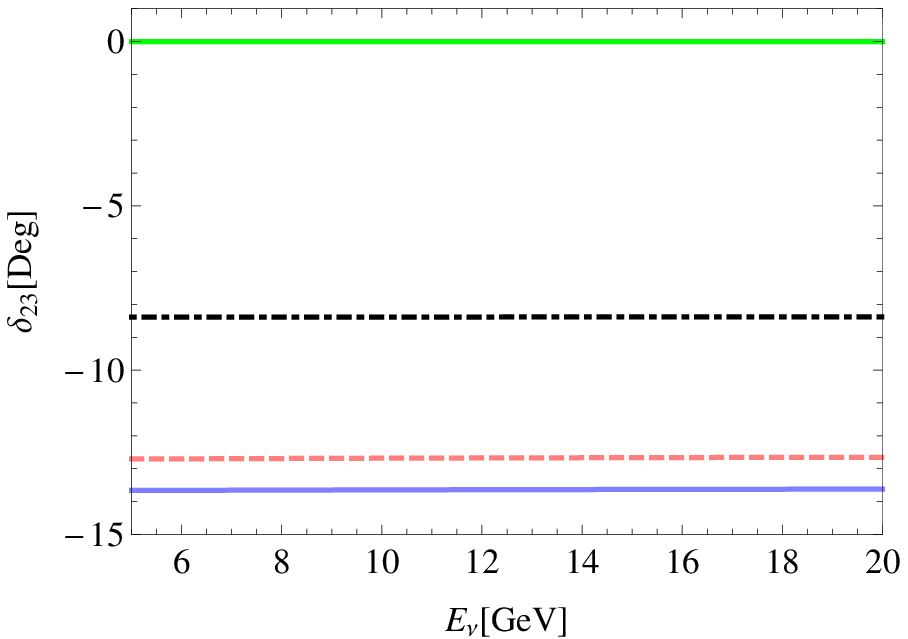}
\caption{DIS ($W'$): The left (right) panel figures illustrate the deviation $\delta_{23}$ with the $W^\prime$ mass $M_{W'}$ ($E_\nu$) when both left and right-handed  $W^\prime$ couplings are present. The lines show predictions for some representative values of the $W^\prime$ couplings $(g^{\tau \nu_\tau}_L, g^{ud}_L, g^{ud}_R)$. The green line (solid, upper) corresponds to the SM prediction. The blue line (solid, lower) in the left  figure corresponds to (-0.94 ,  -1.13 , -0.85) at $E_{\nu} = 17$  GeV, and the blue line (solid, lower) in the right  figure corresponds to (1.23 , 0.84 , 0.61) at $ M_{W^\prime} = 200 $ GeV. Here, we use the best-fit value $  \theta_{13} = 9.1^\circ$ \cite{Tortola:2012te}. }
\label{res-W-delta23-new}
\end{figure}

\begin{figure}[h!]
\centering
 \includegraphics[width=7.25cm]{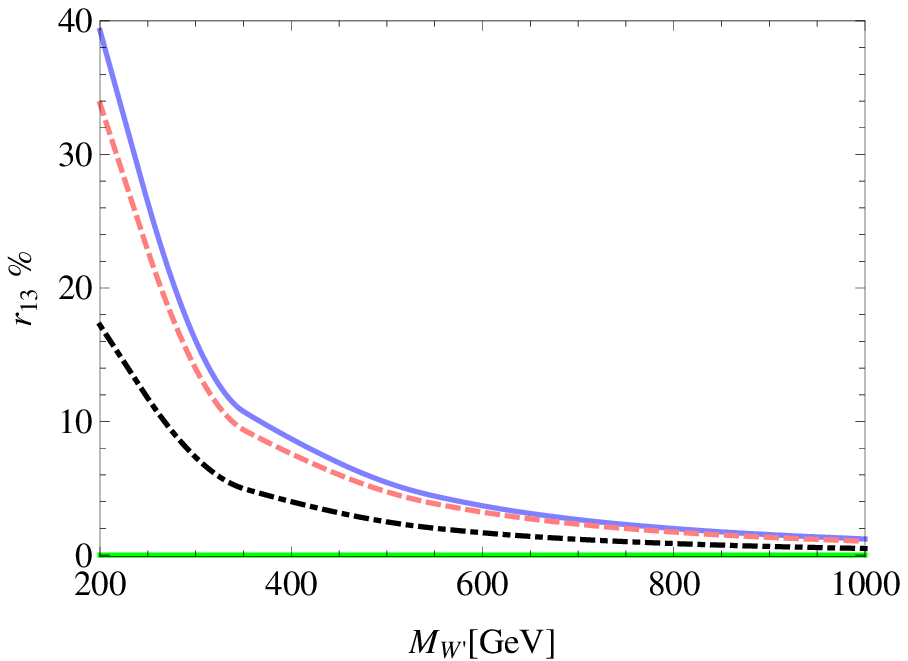}~~~
 \includegraphics[width=7.25cm]{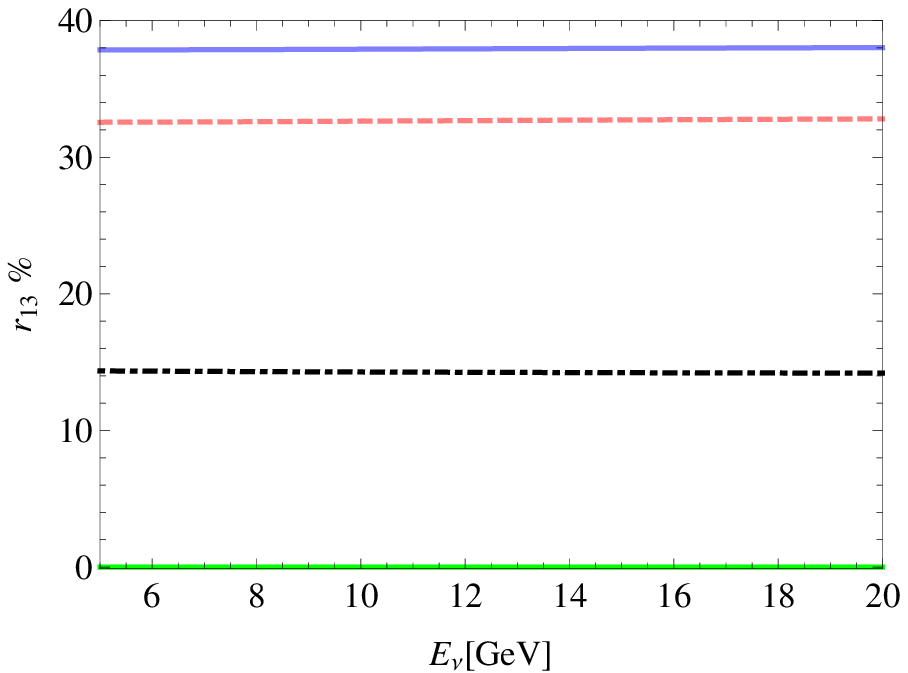}
\caption{DIS ($W'$): The left (right) panel figures illustrate the variation of $r_{W'}^{13}\%$ with the $W^\prime$ mass $M_{W'}$ ($E_\nu$) when both left and right-handed  $W^\prime$ couplings are present. The lines show predictions for some representative values of the $W^\prime$ couplings $(g^{\tau \nu_\tau}_L, g^{ud}_L, g^{ud}_R)$. The green line (solid, lower) corresponds to the SM prediction. The blue line (solid, upper) in the left  figure corresponds to (-0.94 ,  -1.13 , -0.85)  at $E_{\nu} = 17$  GeV, and the blue line (solid, upper) in the right  figure corresponds to (1.23 , 0.84 , 0.61) at $ M_{W^\prime} = 200 $ GeV.}
\label{res-W-r13-new}
\end{figure}

\begin{figure}[h!]
\centering
 \includegraphics[width=7.25cm]{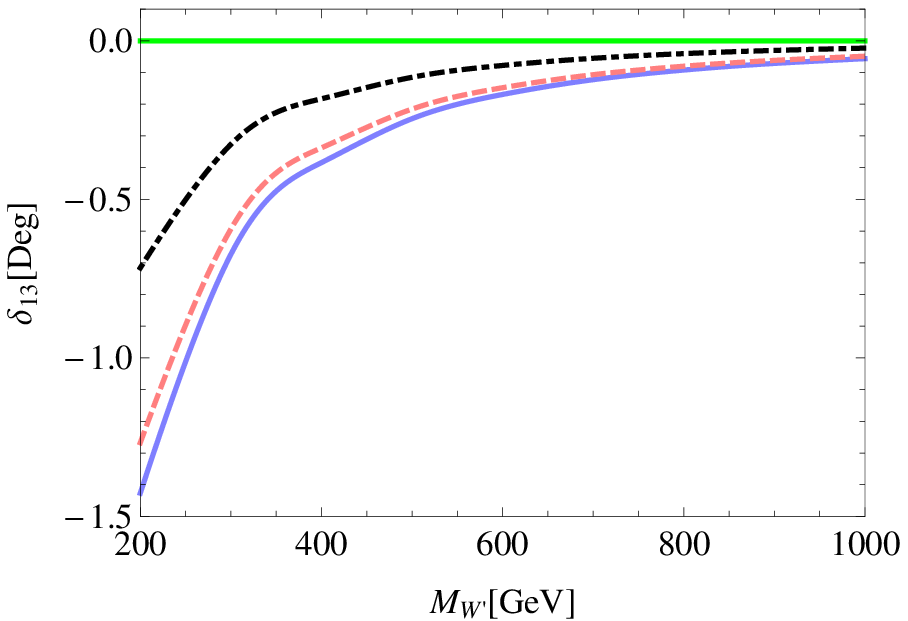}~~~
 \includegraphics[width=7.25cm]{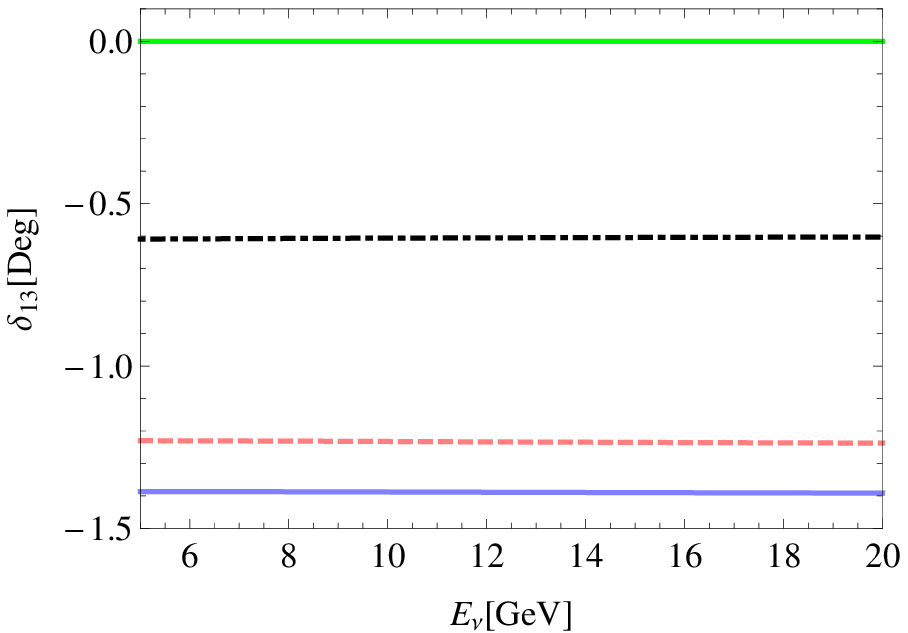}
\caption{DIS ($W'$): The left (right) panel figures illustrate the deviation $\delta_{13}$ with the $W^\prime$ mass $M_{W'}$ ($E_\nu$) when both left and right-handed  $W^\prime$ couplings are present. The lines show predictions for some representative values of the $W^\prime$ couplings $(g^{\tau \nu_\tau}_L, g^{ud}_L, g^{ud}_R)$. The green line (solid, upper) corresponds to the SM prediction. The blue line (solid, lower) in the left  figure corresponds to (-0.94 ,  -1.13 , -0.85) at $E_{\nu} = 17$  GeV, and the blue line (solid, lower) in the right  figure corresponds to (1.23 , 0.84 , 0.61) at $ M_{W^\prime} = 200 $ GeV. Here, we use the best-fit value $  \theta_{13} = 9.1^\circ$ \cite{Tortola:2012te}. }
\label{res-W-delta13-new}
\end{figure}

\begin{figure}[h!]
\centering
 \includegraphics[width=7.25cm]{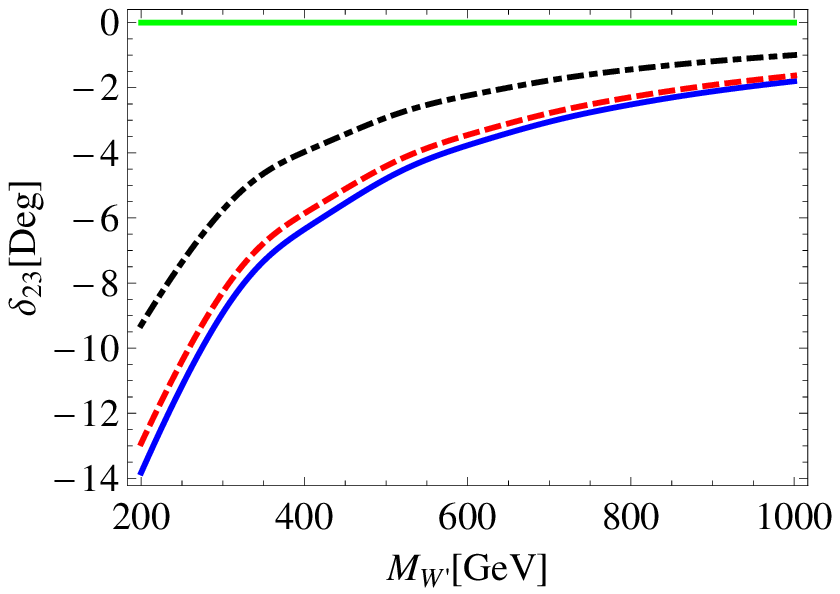}
\caption{DIS ($W'$): The figure illustrates the deviation $\delta_{23}$ with the $W^\prime$ mass $M_{W'}$ when both left and right-handed  $W^\prime$ couplings are present. The lines show predictions for some representative values of the $W^\prime$ couplings $(g^{\tau \nu_\tau}_L, g^{ud}_L, g^{ud}_R)$. The green line (solid, upper) corresponds to the SM prediction. The blue line (solid, lower) corresponds to (-0.94 ,  -1.13 , -0.85). Here, we use the best-fit value $  \theta_{23} = 42.8^\circ$ \cite{GonzalezGarcia:2010er}. We take into account the atmospheric neutrino flux for Kamioka where the Super-Kamiokande experiment locates \cite{Honda:2011nf}.}
\label{DIS-Flux-Wp}
\end{figure}



\section{Polarization of the produced $\tau^\pm$}

In this section we study the effects of NP on the polarization of the produced $\tau$. The starting point is to construct the spin-density matrix $\rho_{\lambda, \lambda^\prime}$, where $\lambda$ and $\lambda^\prime$  are the helicity of the $\tau$ lepton. The spin-density matrix $\rho_{\lambda, \lambda^\prime}$ is related to the spin dependent differential cross section as
\bea
\label{diffgeneq}
\frac{d\sigma_{\lambda, \lambda^\prime}}{d E_l d\cos{\theta}} &=&  |\rho_{\lambda \lambda^\prime}|^2 \frac{d\sigma_{total}}{d E_l d\cos{\theta}} ,
\eea
where the total cross section $\sigma_{total} = \sigma_{\frac{1}{2} \frac{1}{2}} +\sigma_{-\frac{1}{2} -\frac{1}{2}}$.  The spin-density  matrix  $\rho_{\lambda, \lambda^\prime}$ is expressed in terms of the spin dependent matrix element $M_{\lambda, \lambda^\prime} = L^{\mu \nu}_{\lambda, \lambda^\prime} W_{\mu \nu}$ as
\bea
\rho_{\lambda, \lambda^\prime} &=& \frac{{\cal{M}}_{\lambda,\lambda^\prime}} {\sum_{\lambda = \pm \frac{1}{2}}{\cal{M}}_{\lambda, \lambda}}\,.
\eea
The most general form of the polarization density matrix $\rho$ of a fermion is parametrized as
\bea
\label{polvec}
\rho  &=&  [\rho_{\lambda, \lambda^\prime}] = \frac{1}{2}(I + \tau^a \cdot \vec{P}) =  \frac{1}{2} \left(\begin{array}{cc}
 1 + P_z &  P_x - i P_y \\
 P_x + i P_y &  1 - P_z    \\
 \end{array} \right),
\eea
where I is the $2 \times 2$ identity matrix and $\vec{P}$ is the polarization vector of the decaying
spin-1/2 lepton. 

To determine the components $(P_x, P_y, P_z)$ of the  polarization vector we choose the following kinematic variables. The four-momenta of incoming neutrino $(k)$, target nucleon $(p)$ and produced  lepton ($k^\prime$) in the laboratory frame are
\bea
k^\mu &=& (E_\nu, 0, 0, E_\nu)\,, \nl
p^\mu &=& (M, 0, 0, 0)\,, \nl
k^{\prime\mu} &=& (E_l, p_l \sin{\theta} \cos{\phi} , p_l \sin{\theta} \sin{\phi},  p_l \cos{\theta})\,.
\eea
We introduce three four-vectors $s^a_\mu$ , a = 1, 2, 3 such that the $s^a$ and $k'_l/m_l$ form an orthonormal set of four-vectors as defined in  \cite{Haber:1994pe}:
We choose the three spin four-vectors of the lepton such that
\bea
 s^a \cdot k^\prime &=& 0,\nonumber\\
 s^a \cdot s^b &=& - {\delta}^{ab},\nonumber\\
 s_\mu^a \cdot s_\nu^b &=& -g_{\mu\nu}+\frac{k'_\mu k'_\nu}{m_l^2},
\eea
where
\bea
s^1_\mu &=& (0, \cos{\theta} \cos{\phi}, \cos{\theta} \sin{\phi}, - \sin{\theta}) \,, \nl
s^2_\mu &=& (0, -\sin{\phi}, \cos{\phi}, 0) \,, \nl
s^3_\mu &=& (p_l/m_l, E_l/m_l \sin{\theta} \cos{\phi}, E_l/m_l \sin{\theta} \sin{\phi}, E_l/m_l \cos{\theta}) \,.
\eea
Finally we define the degree of $\tau$ polarization P as
\bea
\label{Pdef}
P = \sqrt{P^2_x + P^2_y + P^2_z}.
\eea
The SM results for the polarization components $P_x,P_y, P_z$ can be found in Ref.~\cite{Hagiwara:2003di} for the processes  QE, $\Delta$-RES and DIS. We calculated these components in the presence of the charged Higgs and $W^\prime$ contributions. 
We computed the degree of $\tau$ polarization $P$ with respect to  $E_\tau$  for 0 degree, 5 degrees and 10 degrees scattering angles with the incident neutrino energy  at 10 GeV. In the polarization results we found the charged Higgs and $W'$ model produce tiny deviations from the SM values.



\section{Conclusion}

New physics contributions to the tau-neutrino nucleon scattering were considered in this work. We discussed charged Higgs and $W'$ effects to the $\Delta$ resonance production $\scatdeltamix$ and deep inelastic scattering $\scatmix$ in the neutrino-nucleon interactions. Considering these effects in the neutrino detection process at neutrino oscillation experiments modify the measured atmospheric and reactor mixing angles $\theta_{23}$ and $\theta_{13}$, respectively. In the resonance production, we included form factor effects in the calculations of the deviations $\delta_{23}$ and $\delta_{13}$ of the actual mixing angles from the measured ones. We constrained the parameters of both  models from $\taud$ and $\tauv$ decays that is discussed in a previous work. The cross section of the $\Delta$ resonance production was calculated within the kinematical interval $M + m_\pi < W < W_{cut}$, while the deep inelastic scattering was calculated within the range $W_{cut} 
< W < \sqrt{s}-m_\tau$  with $W_{cut}=1.4$ GeV. If high-energy LBL experiments could measure $\theta_{13}$ via $\nu_\tau$ appearance, the NP effects can impact the $\theta_{13}$ measurement. As $\theta_{13}$ is a small angle, large NP parameters are required to produce  observable deviations $\delta_{13}$. 

In the case of $\Delta$ resonance production, the charged Higgs contribution was found to be proportional to $q^2$ which suppressed the NP effect within the allowed kinematical region. The values of the deviations $\delta_{23}$ and $\delta_{13}$  were negative as the interference term in the cross section vanishes in the limit of ignoring the neutrino mass and, in turn, the total cross section is always larger than the SM one. The values of $\delta_{23}$ and $\delta_{13}$ in the $W'$ gauge boson contributions were found to be both positive and negative, but were mostly negative. The $\delta_{23}$ and $\delta_{13}$ values decreased in magnitude with increasing  incident neutrino energy and the new state masses $(M_{W'},M_H)$.

In the case of deep inelastic scattering, the charged Higgs contribution does not have interference with the SM cross section. With the constraints on the NP parameters, the NP effects were negligible and the deviations $\delta_{23}$ and $\delta_{13}$ were very small. The values of deviations were found to be mostly negative in the $W'$ model. The $\delta_{23}$ and $\delta_{13}$ values increased in magnitude with increasing incident neutrino energy and decreased with increasing $M_{W'}$. 

We took into account the flux of incoming atmospheric neutrinos from Kamioka, where the Super-Kamiokande experiment is located, in the calculations of $\delta_{23}$ when considering the charged Higgs and $W'$ contributions. By integrating over the incoming neutrino energy we found that considering the neutrino flux did not change the $\delta_{23}$ results significantly. 
We showed the $ 3 \sigma$ and $5 \sigma$ deviation contour plots, using the $\chi^2$ measure and the $W^\prime$ NSI model, for the number of events  for neutrino energies above $30$ GeV where the DIS contribution is dominant
Finally, we studied the NP effects on the degree of polarization of the produced $\tau$ and found that the deviation of the polarization results in the NP models  from the SM values  were negligibly small at different  scattering angles.

\section*{Acknowledgements} We thank Sandip Pakvasa, Nita Sinha, and Murugeswaran Duraisamy for useful comments and discussions. This work was financially supported in part by the National Science Foundation under Grant No.\ NSF PHY-1068052.


\clearpage


\begin{thebibliography}{99}




\bibitem{Wolfenstein:1977ue}
L.~Wolfenstein, Phys. Rev.D17, 2369, 1978.

\bibitem{Mikheev:1986gs}
S.~P. Mikheyev and A.~Y. Smirnov,
Sov. J. Nucl. Phys. 42, 913, 1985.



\bibitem{Roulet:1991sm} 
  E.~Roulet,
  Phys.\ Rev.\ D {\bf 44}, 935 (1991).

\bibitem{Brooijmans:1998py} 
  G.~Brooijmans,
  hep-ph/9808498.

\bibitem{GonzalezGarcia:1998hj} 
  M.~C.~Gonzalez-Garcia, M.~M.~Guzzo, P.~I.~Krastev, H.~Nunokawa, O.~L.~G.~Peres, V.~Pleitez, J.~W.~F.~Valle and R.~Zukanovich Funchal,
  Phys.\ Rev.\ Lett.\  {\bf 82}, 3202 (1999)
  [hep-ph/9809531].


\bibitem{Guzzo:1991hi} 
  M.~M.~Guzzo, A.~Masiero and S.~T.~Petcov,
  Phys.\ Lett.\ B {\bf 260}, 154 (1991).


\bibitem{Bergmann:2000gp} 
  S.~Bergmann, M.~M.~Guzzo, P.~C.~de Holanda, P.~I.~Krastev and H.~Nunokawa,
  Phys.\ Rev.\ D {\bf 62}, 073001 (2000)
  [hep-ph/0004049].

\bibitem{Guzzo:2000kx} 
  M.~M.~Guzzo, H.~Nunokawa, P.~C.~de Holanda and O.~L.~G.~Peres,
  Phys.\ Rev.\ D {\bf 64}, 097301 (2001)
  [hep-ph/0012089].

\bibitem{Guzzo:2001mi} 
  M.~Guzzo, P.~C.~de Holanda, M.~Maltoni, H.~Nunokawa, M.~A.~Tortola and J.~W.~F.~Valle,
  Nucl.\ Phys.\ B {\bf 629}, 479 (2002)
  [hep-ph/0112310].


\bibitem{Grossman:1995wx} 
  Y.~Grossman,
  Phys.\ Lett.\ B {\bf 359}, 141 (1995)
  [hep-ph/9507344].


\bibitem{Ota:2002na} 
  T.~Ota and J.~Sato,
  Phys.\ Lett.\ B {\bf 545}, 367 (2002)
  [hep-ph/0202145].

\bibitem{Friedland:2005vy} 
  A.~Friedland and C.~Lunardini,
  Phys.\ Rev.\ D {\bf 72}, 053009 (2005)
  [hep-ph/0506143].





\bibitem{Kitazawa:2006iq} 
  N.~Kitazawa, H.~Sugiyama and O.~Yasuda,
  hep-ph/0606013.


\bibitem{Friedland:2006pi} 
  A.~Friedland and C.~Lunardini,
  Phys.\ Rev.\ D {\bf 74}, 033012 (2006)
  [hep-ph/0606101].


\bibitem{Blennow:2007pu} 
  M.~Blennow, T.~Ohlsson and J.~Skrotzki,
  Phys.\ Lett.\ B {\bf 660}, 522 (2008)
  [hep-ph/0702059 [HEP-PH]].

\bibitem{EstebanPretel:2008qi} 
  A.~Esteban-Pretel, J.~W.~F.~Valle and P.~Huber,
  Phys.\ Lett.\ B {\bf 668}, 197 (2008)
  [arXiv:0803.1790 [hep-ph]].

\bibitem{Blennow:2008ym} 
  M.~Blennow, D.~Meloni, T.~Ohlsson, F.~Terranova and M.~Westerberg,
  Eur.\ Phys.\ J.\ C {\bf 56}, 529 (2008)
  [arXiv:0804.2744 [hep-ph]].

\bibitem{GonzalezGarcia:2001mp} 
  M.~C.~Gonzalez-Garcia, Y.~Grossman, A.~Gusso and Y.~Nir,
  Phys.\ Rev.\ D {\bf 64}, 096006 (2001)
  [hep-ph/0105159].


\bibitem{Gago:2001xg} 
  A.~M.~Gago, M.~M.~Guzzo, H.~Nunokawa, W.~J.~C.~Teves and R.~Zukanovich Funchal,
  Phys.\ Rev.\ D {\bf 64}, 073003 (2001)
  [hep-ph/0105196].



\bibitem{Huber:2001zw} 
  P.~Huber and J.~W.~F.~Valle,
  Phys.\ Lett.\ B {\bf 523}, 151 (2001)
  [hep-ph/0108193].




\bibitem{Ota:2001pw} 
  T.~Ota, J.~Sato and N.~-a.~Yamashita,
  Phys.\ Rev.\ D {\bf 65}, 093015 (2002)
  [hep-ph/0112329].

\bibitem{Campanelli:2002cc} 
  M.~Campanelli and A.~Romanino,
  Phys.\ Rev.\ D {\bf 66}, 113001 (2002)
  [hep-ph/0207350].

\bibitem{Blennow:2005qj} 
  M.~Blennow, T.~Ohlsson and W.~Winter,
  Eur.\ Phys.\ J.\ C {\bf 49}, 1023 (2007)
  [hep-ph/0508175].

\bibitem{Kopp:2007mi} 
  J.~Kopp, M.~Lindner and T.~Ota,
  Phys.\ Rev.\ D {\bf 76}, 013001 (2007)
  [hep-ph/0702269 [HEP-PH]].

\bibitem{Kopp:2007ne} 
  J.~Kopp, M.~Lindner, T.~Ota and J.~Sato,
  Phys.\ Rev.\ D {\bf 77}, 013007 (2008)
  [arXiv:0708.0152 [hep-ph]].

\bibitem{Ribeiro:2007ud} 
  N.~C.~Ribeiro, H.~Minakata, H.~Nunokawa, S.~Uchinami and R.~Zukanovich-Funchal,
  JHEP {\bf 0712}, 002 (2007)
  [arXiv:0709.1980 [hep-ph]].


\bibitem{Bandyopadhyay:2007kx} 
  A.~Bandyopadhyay {\it et al.}  [ISS Physics Working Group Collaboration],
  Rept.\ Prog.\ Phys.\  {\bf 72}, 106201 (2009)
  [arXiv:0710.4947 [hep-ph]].

\bibitem{Ribeiro:2007jq} 
  N.~C.~Ribeiro, H.~Nunokawa, T.~Kajita, S.~Nakayama, P.~Ko and H.~Minakata,
  Phys.\ Rev.\ D {\bf 77}, 073007 (2008)
  [arXiv:0712.4314 [hep-ph]].

\bibitem{Kopp:2008ds} 
  J.~Kopp, T.~Ota and W.~Winter,
  Phys.\ Rev.\ D {\bf 78}, 053007 (2008)
  [arXiv:0804.2261 [hep-ph]].

\bibitem{Malinsky:2008qn} 
  M.~Malinsky, T.~Ohlsson and H.~Zhang,
  Phys.\ Rev.\ D {\bf 79}, 011301 (2009)
  [arXiv:0811.3346 [hep-ph]].

\bibitem{Gago:2009ij} 
  A.~M.~Gago, H.~Minakata, H.~Nunokawa, S.~Uchinami and R.~Zukanovich Funchal,
  JHEP {\bf 1001}, 049 (2010)
  [arXiv:0904.3360 [hep-ph]].



\bibitem{Palazzo:2009rb} 
  A.~Palazzo and J.~W.~F.~Valle,
  Phys.\ Rev.\ D {\bf 80}, 091301 (2009)
  [arXiv:0909.1535 [hep-ph]].
    A.~Palazzo,
    Phys.\ Rev.\ D {\bf 83}, 101701 (2011)
    [arXiv:1101.3875 [hep-ph]].



\bibitem{Meloni:2009cg} 
  D.~Meloni, T.~Ohlsson, W.~Winter and H.~Zhang,
  JHEP {\bf 1004}, 041 (2010)
  [arXiv:0912.2735 [hep-ph]].


\bibitem{Coloma:2011rq} 
  P.~Coloma, A.~Donini, J.~Lopez-Pavon and H.~Minakata,
  JHEP {\bf 1108}, 036 (2011)
  [arXiv:1105.5936 [hep-ph]].



\bibitem{Mitsuka:2011ty} 
  G.~Mitsuka {\it et al.}  [Super-Kamiokande Collaboration],
  Phys.\ Rev.\ D {\bf 84}, 113008 (2011)
  [arXiv:1109.1889 [hep-ex]].



\bibitem{Adhikari:2012vc} 
  R.~Adhikari, S.~Chakraborty, A.~Dasgupta and S.~Roy,
  Phys.\ Rev.\ D {\bf 86}, 073010 (2012)
  [arXiv:1201.3047 [hep-ph]].


\bibitem{Agarwalla:2012wf} 
  S.~K.~Agarwalla, F.~Lombardi and T.~Takeuchi,
  JHEP {\bf 1212}, 079 (2012)
  [arXiv:1207.3492 [hep-ph]].



\bibitem{Ohlsson:2013epa} 
  T.~Ohlsson, H.~Zhang and S.~Zhou,
  Phys.\ Rev.\ D {\bf 88}, 013001 (2013)
  [arXiv:1303.6130 [hep-ph]].







\bibitem{davidson} 
S.~Davidson, C.~Pena-Garay, N.~Rius and A.~Santamaria,
JHEP {\bf 0303}, 011 (2003)
[arXiv:hep-ph/0302093].


\bibitem{DELPHI} 
J.~Abdallah {\it et al.} [DELPHI Collaboration],
Eur.\ Phys.\ J.\ C {\bf 38}, 395 (2005)
[arXiv:hep-ex/0406019].

\bibitem{Biggio:2009nt} 
  C.~Biggio, M.~Blennow and E.~Fernandez-Martinez,
  JHEP {\bf 0908}, 090 (2009)
  [arXiv:0907.0097 [hep-ph]].
 
  
\bibitem{Berezhiani:2001rt} 
  Z.~Berezhiani, R.~S.~Raghavan and A.~Rossi,
  Nucl.\ Phys.\ B {\bf 638}, 62 (2002)
  [hep-ph/0111138].
  
  
  \bibitem{Friedland:2004pp} 
    A.~Friedland, C.~Lunardini and C.~Pena-Garay,
    Phys.\ Lett.\ B {\bf 594}, 347 (2004)
    [hep-ph/0402266].
    
    
    
\bibitem{Miranda:2004nb} 
  O.~G.~Miranda, M.~A.~Tortola and J.~W.~F.~Valle,
  JHEP {\bf 0610}, 008 (2006)
  [hep-ph/0406280].
  
  

\bibitem{Bergmann:1999pk} 
  S.~Bergmann, Y.~Grossman and D.~M.~Pierce,
  Phys.\ Rev.\ D {\bf 61}, 053005 (2000)
  [hep-ph/9909390].
  
  
\bibitem{Fornengo:2001pm} 
  N.~Fornengo, M.~Maltoni, R.~Tomas and J.~W.~F.~Valle,
  Phys.\ Rev.\ D {\bf 65}, 013010 (2002)
  [hep-ph/0108043].
  
  
  \bibitem{GonzalezGarcia:2004wg} 
    M.~C.~Gonzalez-Garcia and M.~Maltoni,
    Phys.\ Rev.\ D {\bf 70}, 033010 (2004)
    [hep-ph/0404085].


\bibitem{Friedland:2004ah} 
  A.~Friedland, C.~Lunardini and M.~Maltoni,
  Phys.\ Rev.\ D {\bf 70}, 111301 (2004)
  [hep-ph/0408264].

\bibitem{Escrihuela:2009up} 
  F.~J.~Escrihuela, O.~G.~Miranda, M.~A.~Tortola and J.~W.~F.~Valle,
  Phys.\ Rev.\ D {\bf 80}, 105009 (2009)
  [Erratum-ibid.\ D {\bf 80}, 129908 (2009)]
  [arXiv:0907.2630 [hep-ph]].
  
\bibitem{Barranco:2008rc} 
  J.~Barranco, O.~G.~Miranda and T.~I.~Rashba,
  Nucl.\ Phys.\ Proc.\ Suppl.\  {\bf 188}, 214 (2009)
  [arXiv:0810.5361 [hep-ph]].
  





\bibitem{Barranco:2005yy} 
  J.~Barranco, O.~G.~Miranda and T.~I.~Rashba,
  JHEP {\bf 0512}, 021 (2005)
  [hep-ph/0508299].
  \bibitem{Barranco:2007tz} 
    J.~Barranco, O.~G.~Miranda and T.~I.~Rashba,
    Phys.\ Rev.\ D {\bf 76}, 073008 (2007)
    [hep-ph/0702175].






\bibitem{chwprime}
See for example R.~A.~Diaz,
  hep-ph/0212237;
A.~Datta, P.~J.~O'Donnell, Z.~H.~Lin, X.~Zhang and T.~Huang,
  Phys.\ Lett.\ B {\bf 483}, 203 (2000)
  [hep-ph/0001059].


\bibitem{Abe:2012jj} 
  K.~Abe {\it et al.}  [Super-Kamiokande Collaboration],
  arXiv:1206.0328 [hep-ex].

  \bibitem{Abe:2006fu} 
    K.~Abe {\it et al.}  [Super-Kamiokande Collaboration],
    Phys.\ Rev.\ Lett.\  {\bf 97}, 171801 (2006)
    [hep-ex/0607059].




\bibitem{:2011ph} 
  [OPERA Collaboration],
  arXiv:1107.2594 [hep-ex].
  B.~Wonsak [OPERA Collaboration],
  J.\ Phys.\ Conf.\ Ser.\  {\bf 335}, 012051 (2011).




%
%
%
%
%







\bibitem{Kodama:2007aa} 
  K.~Kodama {\it et al.}  [DONuT Collaboration],
  Phys.\ Rev.\ D {\bf 78}, 052002 (2008)
  [arXiv:0711.0728 [hep-ex]].


\bibitem{Rashed:2012bd} 
  A.~Rashed, M.~Duraisamy and A.~Datta,
  arXiv:1204.2023 [hep-ph].

\bibitem{Jadach:1993hs} 
  S.~Jadach, Z.~Was, R.~Decker and J.~H.~Kuhn,
  Comput.\ Phys.\ Commun.\  {\bf 76}, 361 (1993).



\bibitem{Hagiwara:2003di} 
 K.~Hagiwara, K.~Mawatari and H.~Yokoya,
 Nucl.\ Phys.\ B {\bf 668}, 364 (2003)
 [Erratum-ibid.\ B {\bf 701}, 405 (2004)]
 [hep-ph/0305324].




\bibitem{py}
E. A. Paschos and J. Y. Yu, Phys. Rev. D{\bf 65}(2002)033002.



\bibitem{sehgal}
D. Rein and L. M. Sehgal, Ann. Phys. {\bf 133}(1981)79.

\bibitem{ppy}
E. A. Paschos, L. Pasquali and J. Y. Yu, Nucl. Phys. B{\bf 588}(2000)263.


\bibitem{kretzer}
S. Kretzer and M. H. Reno, Phys. Rev. D{\bf 66}(2002)113007.

\bibitem{albright}
C. H. Albright and C. Jarlskog, Nucl. Phys. B{\bf 84}(1975)467.



\bibitem{smith}
C. H. Llewellyn Smith, Phys. Rep. {\bf 3}(1972)261.

\bibitem{hippel}
P. A. Schreiner and  F. Von Hippel, Nucl. Phys. B{\bf 58}(1973)333.

\bibitem{fogli}
G. L. Fogli and G. Nardulli, Nucl. Phys. B{\bf 160}(1979)116.

\bibitem{alvarez}
L. Alvarez-Ruso, S. K. Singh and M. J. Vicente Vacas,
Phys. Rev. C{\bf 57}(1998)2693.


\bibitem{singh}
S. K. Singh, Nucl. Phys. B (Proc. Suppl.) {\bf 112}(2002)77.


\bibitem{Diaz:2002tp} 
 See, e.g., R.~A.~Diaz,
  hep-ph/0212237, 
    O.~Deschamps, S.~Descotes-Genon, S.~Monteil, V.~Niess, S.~T'Jampens and V.~Tisserand,
    Phys.\ Rev.\ D {\bf 82}, 073012 (2010)
    [arXiv:0907.5135 [hep-ph]],
    G.~C.~Branco, P.~M.~Ferreira, L.~Lavoura, M.~N.~Rebelo, M.~Sher and J.~P.~Silva,
    arXiv:1106.0034 [hep-ph].







\bibitem{Honda:2011nf} 
  M.~Honda, T.~Kajita, K.~Kasahara and S.~Midorikawa,
  Phys.\ Rev.\ D {\bf 83}, 123001 (2011)
  [arXiv:1102.2688 [astro-ph.HE]].


\bibitem{PDG-No}
W.M. Yao et al. (Particle Data Group), J. Phys. G 33, 1 (2006).


\bibitem{Adamson:2009ju} 
  P.~Adamson {\it et al.}  [MINOS Collaboration],
  Phys.\ Rev.\ D {\bf 81}, 072002 (2010)
  [arXiv:0910.2201 [hep-ex]].



\bibitem{Honda:2004yz} 
  M.~Honda, T.~Kajita, K.~Kasahara and S.~Midorikawa,
  Phys.\ Rev.\ D {\bf 70}, 043008 (2004)
  [astro-ph/0404457].


\bibitem{Conrad:2010mh} 
  J.~Conrad, A.~de Gouvea, S.~Shalgar and J.~Spitz,
  Phys.\ Rev.\ D {\bf 82}, 093012 (2010)
  [arXiv:1008.2984 [hep-ph]].

\bibitem{GonzalezGarcia:2010er} 
  M.~C.~Gonzalez-Garcia, M.~Maltoni and J.~Salvado,
  JHEP {\bf 1004}, 056 (2010)
  [arXiv:1001.4524 [hep-ph]].




\bibitem{Tortola:2012te} 
  D.~V.~Forero, M.~Tortola and J.~W.~F.~Valle,
  arXiv:1205.4018 [hep-ph].


\bibitem{Haber:1994pe} 
  H.~E.~Haber,
  In *Stanford 1993, Spin structure in high energy processes* 231-272
  [hep-ph/9405376].



\end{thebibliography}
\end{document}